\documentclass[reprint]{JASA}

\usepackage[acronyms, nonumberlist, nogroupskip, automake]{glossaries}
\usepackage{lmodern}
\usepackage[]{siunitx}
\usepackage{upgreek}
\usepackage{subfigure}
\usepackage{placeins}
\usepackage{cleveref}
\Crefname{figure}{Fig.}{Figs.}
\Crefname{tabular}{Tab.}{Tabs.}
\Crefname{section}{Sec.}{Secs.}

\DeclareMathOperator{\sinc}{sinc}

\newcommand{\subfigwidth}{\ifdim\columnwidth=\textwidth 225pt \else \columnwidth \fi}

\makeglossaries

\newacronym{6dof}{6DOF}{Six Degrees of Freedom}
\newacronym{aa}{AA}{Acoustic Acquisition}
\newacronym{ahc}{AHC}{Agglomerative Hierarchical Clustering}
\newacronym{ail}{AIL}{Alamire Interactive Lab}
\newacronym{allrad}{ALLRAD}{All-Round Ambisonic Decoding}
\newacronym{art}{ART}{Acoustic Ray Tracing}
\newacronym{asw}{ASW}{Apparent Source Width}
\newacronym{cad}{CAD}{Computer Aided Design}
\newacronym{csd}{CSD}{Cross-Spectral Density}
\newacronym{cvp}{CVP}{Consensus Vocabulary Profiling}
\newacronym{dirac}{DirAC}{Directional Audio Coding}
\newacronym{dmfa}{DMFA}{Dual Multiple Factor Analysis}
\newacronym{doa}{DOA}{Direction of Arrival}
\newacronym{dof}{DOF}{Degrees of Freedom}
\newacronym{drr}{DRR}{Direct-to-Reverberant Ratio}
\newacronym{dtft}{DTFT}{Discrete-Time Fourier Transform}
\newacronym{edt}{EDT}{Early Decay Time}
\newacronym{ess}{ESS}{Exponential Sine Sweep}
\newacronym{fcp}{FCP}{Free Choice Profiling}
\newacronym{fcr}{FCR}{Feedback Canceling Reverberator}
\newacronym{fdn}{FDN}{Feedback Delay Network}
\newacronym{fft}{FFT}{Fast Fourier Transform}
\newacronym{fir}{FIR}{Finite Impulse Response}
\newacronym{foa}{FOA}{First-Order Ambisonics}
\newacronym{fp}{FP}{Flash Profile}
\newacronym{fwht}{FWHT}{Fast Walsh-Hadamard Transform}
\newacronym{gcc}{GCC}{Generalized Cross-Correlation}
\newacronym{glram}{GLRAM}{Generalized Low-Rank Approximation of Matrices}
\newacronym{gpa}{GPA}{Generalized Procrustes Analysis}
\newacronym{gpu}{GPU}{Graphics Processing Unit}
\newacronym{hoa}{HOA}{Higher-Order Ambisonic}
\newacronym{hrtf}{HRTF}{Head-Related Transfer Function}
\newacronym{hws}{HWS}{Historical Worship Spaces}
\newacronym{icc}{ICC}{Interaural Cross-Correlation}
\newacronym{iir}{IIR}{Infinite Impulse Response}
\newacronym{imu}{IMU}{Inertial Measurement Unit}
\newacronym{ir}{IR}{Impulse Response}
\newacronym{ism}{ISM}{Image-Source Method}
\newacronym{ivp}{IVP}{Individual Vocabulary Profiling}
\newacronym{jnd}{JND}{Just Noticeable Difference}
\newacronym{lti}{LTI}{Linear Time-Invariant}
\newacronym{md}{MD}{Molecular Dynamics}
\newacronym{mdap}{MDAP}{Multiple-Direction Amplitude Panning}
\newacronym{mfa}{MFA}{Multiple Factor Analysis}
\newacronym{ml}{ML}{Maximum-Likelihood}
\newacronym{mls}{MLS}{Maximum Length Sequence}
\newacronym{mushra}{MUSHRA}{MUltiple Stimuli with Hidden Reference and Anchor}
\newacronym{nls}{NLS}{Nearest Loudspeaker Synthesis}
\newacronym{nmse}{NMSE}{Normalized Mean Square Error}
\newacronym{nupc}{NUPC}{Non-Uniform Partitioned Convolution}
\newacronym{ola}{OLA}{Overlap-Add}
\newacronym{ols}{OLS}{Overlap-Save}
\newacronym{orap}{ORAP}{Objective Room Acoustic Parameters}
\newacronym{pca}{PCA}{Principal Component Analysis}
\newacronym{pe}{PE}{Perceptual Evaluation}
\newacronym{qda}{QDA}{Quantitative Descriptive Analysis}
\newacronym{rir}{RIR}{Room Impulse Response}
\newacronym{rt}{RT}{Reverberation Time}
\newacronym{rta}{RTA}{Real-Time Auralization}
\newacronym{sdm}{SDM}{Spatial Decomposition Method}
\newacronym{sdn}{SDN}{Scattering Delay Network}
\newacronym{sfa}{SFA}{Sound Field Analysis}
\newacronym{sfs}{SFS}{Sound Field Synthesis}
\newacronym{shd}{SHD}{Spherical Harmonic Decomposition}
\newacronym{snr}{SNR}{Signal-to-Noise Ratio}
\newacronym{srir}{SRIR}{Spatial Room Impulse Response}
\newacronym{ssw}{SSW}{SpatialSound Wave}
\newacronym{sti}{STI}{Speech Transmission Index}
\newacronym{sv}{SV}{Slowness Vector}
\newacronym{svd}{SVD}{Singular Value Decomposition}
\newacronym{tde}{TDE}{Time Delay Estimation}
\newacronym{tdoa}{TDOA}{Time Difference of Arrival}
\newacronym{toa}{TOA}{Time of Arrival}
\newacronym{vbap}{VBAP}{Vector Base Amplitude Panning}
\newacronym{vr}{VR}{Virtual Reality}
\newacronym{wfa}{WFA}{Wave Field Analysis}
\newacronym{wfs}{WFS}{Wave Field Synthesis}
 
\begin{document}

\title[]{On Time Delay Interpolation for Improved Acoustic Reflector Localization}

\author{Hannes Rosseel}
\author{Toon van Waterschoot}

\affiliation{KU Leuven, Dept. of Electrical Engineering (ESAT), STADIUS Center for Dynamical Systems, Signal Processing and Data Analytics, Leuven, Belgium}

\email{hannes.rosseel@esat.kuleuven.be}

\date{} 
\glsdisablehyper

\begin{abstract}
\vspace{2pt}
The localization of acoustic reflectors is a fundamental component in various applications, including room acoustics analysis, sound source localization, and acoustic scene analysis. \gls{tde} is essential for determining the position of reflectors relative to a sensor array. Traditional \gls{tde} algorithms generally yield time delays that are integer multiples of the operating sampling period, potentially lacking sufficient time resolution. To achieve subsample \gls{tde} accuracy, various interpolation methods, including parabolic, Gaussian, frequency, and sinc interpolation, have been proposed. This paper presents a comprehensive study on time delay interpolation to achieve subsample accuracy for acoustic reflector localization in reverberant conditions. We derive the Whittaker-Shannon interpolation formula from the previously proposed sinc interpolation in the context of short-time windowed \gls{tde} for acoustic reflector localization. Simulations show that sinc and Whittaker-Shannon interpolation outperform existing methods in terms of time delay error and positional error for critically sampled and band-limited reflections. Performance is evaluated on real-world measurements from the \textit{MYRiAD} dataset, showing that sinc and Whittaker-Shannon interpolation consistently provide reliable performance across different sensor-source pairs and loudspeaker positions. These results can enhance the precision of acoustic reflector localization systems, vital for applications such as room acoustics analysis, sound source localization, and acoustic scene analysis.
\glsresetall \end{abstract}

\maketitle

\section{Introduction}

Accurately estimating the position of acoustic reflectors is of great importance for many acoustical signal processing applications. These applications include inferring room geometry \cite{antonacci2010geometric, antonacci2012inference, tervo2012room, annibale2012tdoa, el2017room}, analyzing and visualizing room acoustics \cite{patynen2013analysis}, and spatial audio encoding for the auralization \cite{tervo2013sdm} of concert halls \cite{lokki2016concert}, listening environments \cite{tervo2014preferences}, and car cabins \cite{kaplanis2017rapid, camilleri2019evaluation}.

Acoustic reflector localization relies on the time delay information between the arrival of an acoustic wavefront at different sensor positions. The position of an acoustic reflector in 3D space can be estimated by measuring the \gls{toa} and \gls{tdoa} of an acoustic wavefront at a set of spatially separated sensors. The \gls{toa} can be estimated using peak-picking methods \cite{defrance2008detecting, defrance2009using,antonacci2012inference, tervo2012acoustic, remaggi2017acoustic, diego2021dechorate} or maximum-likelihood estimation \cite{ehrenberg1978signal}. The \gls{tdoa} information, which is the time difference between the arrival of a wavefront at two spatially separated sensors, is commonly derived using cross-correlation-based methods, such as the \gls{gcc} method \cite{knapp1976generalized}. The \gls{tdoa} corresponds to the time instant at which the cross-correlation of two sensor pressure values is maximized, and can be used to determine the \gls{doa} of the acoustic wavefront propagating along the sensors. The combination of \gls{tdoa} and \gls{toa} information can be used to reliably localize acoustic reflectors in a room \cite{tervo2012acoustic, tervo2013sdm}.

The localization of acoustic reflectors from a set of \glspl{rir}, measured by a set of $N$ spatially separated sensors, can be achieved by estimating the equivalent image source position of each acoustic reflection impinging on the sensor array \cite{allen1979image}. The image sources are localized by dividing the set of measured \glspl{rir} into overlapping short-time windows, and applying a localization method to each short-time window \cite{tervo2013sdm}. The window length should be chosen slightly larger than the time it takes for an acoustic reflection to propagate along the sensor array \cite{merimaa2005spatial,pulkki2006spatial,tervo2012acoustic, tervo2013sdm}. It is assumed that only one acoustic reflection $k$ is active in each short-time window. This assumption is generally valid for the direct and early acoustic reflections in the measured \gls{rir} if the dimensions of the sensor array are relatively small compared to the dimensions of the room and the sensor array is not positioned near walls or reflective surfaces.

The precision of acoustic source localization is directly impacted by the accuracy of \gls{tdoa} and \gls{toa} estimation. The temporal resolution of \gls{tdoa} and \gls{toa} estimates for individual acoustic reflections is limited by the sampling rates at which measured \glspl{rir} are sampled. Estimation errors arise when the actual time delay between two acoustic wavefronts arriving at spatially separated sensors is not an integer multiple of the sampling period. Inaccurate \gls{tde} may lead to significant localization errors in the estimated position of the acoustic reflector. These errors can occur when the sampling rate is constrained by available hardware, resulting in insufficient temporal resolution to accurately estimate the position of each acoustic reflector. In general, the \gls{tdoa} and \gls{toa} estimate corresponds to the time instant at which a specific \gls{tde} function is maximized. For \gls{tdoa} estimation, this function is the cross-correlation between two acoustic reflections arriving at spatially separated sensors. For \gls{toa} estimation, this function is the filter-matched \glspl{rir} measured by each sensor. Since the \gls{tde} function is discretized in time, the \gls{tde} is limited to the nearest sample value, which may not provide sufficient temporal resolution for accurate localization of acoustic reflectors.

Interpolation methods can improve the accuracy of \gls{tde} by estimating time delays with subsample precision. Previous studies have proposed various interpolation methods that can be categorized into two groups, namely: time-domain interpolation methods and frequency-domain interpolation methods. Time-domain interpolation methods aim to improve the resolution of the \gls{tde} function by fitting a continuous-time function around its peak. Existing interpolation functions are: parabolic functions \cite{lai1999interpolation}, Gaussian functions \cite{zhang2005cross}, or sinc functions \cite{rosseel2021improved}. Another time-domain approach involves reconstructing the continuous-time band-limited \gls{tde} function from the discretized function by applying Whittaker-Shannon interpolation \cite{shannon1949communication}. Frequency-domain interpolation methods achieve subsample \gls{tde} accuracy by fitting a linear function to the weighted phase spectrum of the \gls{tde} function \cite{svilainis2013subsample, svilainis2019review}. In \Cref{sec:tdeinterp2024:interpolation}, we provide a detailed discussion of these interpolation methods and their application to \gls{tde} for acoustic reflector localization.

In \cite{rosseel2021improved}, the authors demonstrated that improving \gls{tde} through sinc interpolation also improves the localization performance of a single acoustic reflector in anechoic conditions. It was shown that sinc interpolation is able to outperform parabolic and Gaussian interpolation in anechoic conditions. However, its performance has not yet been compared to frequency interpolation. Moreover, the authors have not yet verified the effectiveness of applying sinc interpolation for localizing acoustic reflectors in reverberant conditions when a short time window is introduced to isolate each acoustic reflection. This paper seeks to address these limitations by investigating the impact of short-time windowing on the performance of sinc interpolation for \gls{tdoa} estimation, \gls{toa} estimation, and localization of acoustic reflectors. Additionally, this study compares the performance of sinc interpolation to frequency interpolation and proposes a reformulation of sinc interpolation to improve \gls{tde} of acoustic reflections in reverberant conditions.

This paper is structured as follows: \Cref{sec:tdeinterp2024:methods} describes the assumed signal model and outlines the methods used for acoustic reflector localization. \Cref{sec:tdeinterp2024:interpolation} provides a detailed discussion of various interpolation techniques for improving \gls{tde}. The effect of short-time windowing on the accuracy of \gls{tde} is analyzed in \Cref{sec:tdeinterp2024:windowing}. Subsequently, \Cref{sec:tdeinterp2024:sinc_reformulation} presents a novel reformulation of sinc interpolation for \gls{tde} of acoustic reflections in reverberant conditions. A comparison of the performance of different interpolation methods in various simulated scenarios is presented in \Cref{sec:tdeinterp2024:simulations}. In \Cref{sec:tdeinterp2024:measurements}, the performance of the interpolation methods is evaluated on real-world measurements from the \textit{MYRiAD} dataset. Finally, \Cref{sec:tdeinterp2024:conclusion} summarizes the contributions outlined in this paper.
 \section{Preliminaries and Methods}
\label{sec:tdeinterp2024:methods}

\subsection{Room Impulse Response}

The \gls{rir} can be modeled as the sum of all individual acoustic events \cite{tervo2013sdm} originating from a sound wave propagating from a source position $\mathbf{x}$ to a receiver sensor position $\mathbf{r}_n$, such that

\begin{equation}
    \label{eq:tdeinterp2024:rir}
    h_n(t) \triangleq \sum_{k=1}^{K} h_n^{k}(t) + w_n(t),
\end{equation}

\noindent where $t$ represents time, $n$ is the sensor index, $k = 0, \dots, K$ indicates the index of each acoustic event, and $w_n(t)$ is the measurement and model error. 

A \gls{rir} is typically measured by exciting an enclosed space with a known broadband excitation signal and subsequently recording the resulting sound pressure at a receiver position in the space. The \gls{rir} can be extracted from the measured recording and the excitation signal through a deconvolution process. A common excitation signal for measuring \glspl{rir} is the sine sweep \cite{farina2000simultaneous, muller2001transfer,novak2015synchronized}. The measured \gls{rir} is influenced by the characteristics of the source and the sensor that are used in the measurement process. The \gls{rir} measured by a sensor can be modeled as

\begin{equation}
    \tilde{h}_n(t) = h_n(t) * m_n(t),
\end{equation}

\noindent where $n$ is the sensor index, $\tilde{h}_n(t)$ denotes the measured \gls{rir}, and $m_n(t)$ is the combined impulse response, which describes the characteristics of the source and sensor that were used. It is possible to compensate for the characteristics of the measurement equipment by applying a matched filter to the measured \gls{rir} \cite{antonacci2012inference, ehrenberg1978signal}. This matched filter is dependent on the direction of the source relative to the sensor position when the source has a directional radiation patterns. However, in this paper, we will assume that the source behaves as an omnidirectional point source, and the matched filter is independent of the direction of the source. Moreover, it is assumed that the matched filter is known a priori. In practice however, the matched filter can be found by measuring the free-field impulse response between the excitation source and each sensor. Alternatively, if this free-field impulse response is not available, $m_n(t)$ can be estimated from the measured \gls{rir} by isolating the direct-path component using a windowing function \cite{antonacci2012inference, diego2021dechorate}. The \gls{rir} $h_n(t)$ can be found by cross-correlating the matched filter $m_n(t)$ with the measured \gls{rir} $\tilde{h}_n(t)$ as

\begin{equation}
    h_n(t) = \int_{-\infty}^\infty \tilde{h}_n(\tau) m_n(\tau + t) d\tau,
\end{equation}

The definition for a \gls{rir} described in (\ref{eq:tdeinterp2024:rir}) is commonly used in the literature on acoustic source localization \cite{tervo2012acoustic, tervo2012room, tervo2013sdm, remaggi2017acoustic}. In this paper, we specifically consider rooms with rigid walls, allowing the assumption of ideal specular reflections. Consequently, the phase response from the source to each specular reflection picked up by a sensor is considered frequency-independent. Additionally, the sound source is assumed to behave as an omnidirectional point source, exciting the room with an ideal spherical wave pattern. These assumptions allow the use of the image-source model \cite{allen1979image} to represent the \gls{rir} as the sum of individual acoustic reflections, modeled as image sources at positions $\mathbf{x}_k$, such that

\begin{equation}
    \label{eq:tdeinterp2024:sigmodel_acoust_refl}
    h_n^k(t) = \frac{\delta(t - t_n^k)}{4\pi \left\| \mathbf{r}_n - \mathbf{x}_k \right\|_2},
\end{equation}

\noindent where $\delta(\cdot)$ denotes the Dirac delta function, and $t_n^k$ denotes the \gls{toa} or propagation time of a spherical sound wave originating at image-source position $\mathbf{x}_k$ and arriving at sensor position $\mathbf{r}_n$. The \gls{toa} $t_n^k$ is defined as

\begin{equation}
    \label{eq:tdeinterp2024:toa_def}
    t_n^k \triangleq \| \mathbf{r}_n - \mathbf{x}_k \|_2 \, c^{-1},
\end{equation}

\noindent where $c$ denotes the sound propagation speed in air, estimated to be around \SI{343}{\metre\per\second} at a temperature of \SI{20}{\degreeCelsius}.

Up to this point, we have assumed that the \gls{rir} measured at the receiver is not band-limited. In practice however, the measurement equipment that is used exhibits finite bandwidth due to their inherent frequency response characteristics. Additionally, the sampling operation at the receiver position discretizes the \gls{rir} at a fixed sampling interval $T = \frac{1}{f_\mathrm{s}}$, where $f_\mathrm{s}$ is the sampling rate. This sampling operation introduces a bandlimit on the measured signal at the receiver position in accordance with the Nyquist-Shannon sampling theorem \cite{shannon1949communication}. As a result, when a bandlimit equal to the Nyquist frequency $\frac{f_\mathrm{s}}{2}$ is imposed, the individual acoustic reflections described in (\ref{eq:tdeinterp2024:sigmodel_acoust_refl}) can be modeled as

\begin{equation}
    \label{eq:tdeinterp2024:sigmodel_bandlimit_acoust_refl}
    \hat{h}_n^k(t) = \frac{\sinc\left(f_\mathrm{s} \left(t - t_n^k\right)\right)}{4\pi \left\| \mathbf{r}_n - \mathbf{x}_k \right\|_2},
\end{equation}

\noindent where $\sinc(x) = \sin(\pi x) / (\pi x)$ is the normalized sinc function, and $t = \kappa T, \kappa \in \mathbb{Z}$ when the sampling operation is applied.

\subsection{Localization of individual image-sources}
\label{sec:tdeinterp2024:localization}

The choice of localization method depends on the assumption of the sound propagation model and the measurement conditions. Popular methods employ both \gls{toa} and \gls{tdoa} information \cite{tervo2012acoustic} to estimate the position of an acoustic reflection. For now, we will assume that all acoustic reflections can be perfectly separated in time. In \Cref{sec:tdeinterp2024:windowing}, the separation of individual acoustic reflections from a measured \gls{rir} is discussed in more detail.

The \gls{toa} denotes the precise time at which an acoustic reflection arrives at a sensor position $\mathbf{r}_n$, and is defined in (\ref{eq:tdeinterp2024:toa_def}). The \gls{tdoa} is the relative time difference between the arrival of an acoustic reflection originating from an image-source position $\mathbf{x}_k$ at two sensor positions $\{\mathbf{r}_n$, $\mathbf{r}_m\}$. It can be noted that the \gls{tdoa} can be expressed in terms of the \gls{toa} information as

\begin{equation}
    \tau_{m, n}^k = t_n^k - t_m^k = \frac{||\mathbf{r}_n - \mathbf{x}_k||_2 - ||\mathbf{r}_m - \mathbf{x}_k ||_2}{c}.
\end{equation}

The \gls{tdoa} information between two \glspl{rir} is estimated using cross-correlation-based methods, such as the \gls{gcc} \cite{knapp1976generalized}. For an acoustic reflection $k$, measured at two distinct sensor positions $\mathbf{r}_m$ and $\mathbf{r}_n$, the \gls{gcc} $r^k_{m,n}(\tau)$ is defined as:

\begin{equation}
    r^k_{m,n}(\tau) = \int_{-\infty}^{\infty} \Psi(\omega) G^k_{m,n}(\omega) e^{j \omega \tau} d\omega,
\end{equation}

\noindent where $G^k_{m,n}(\omega)$ denotes the \gls{csd} between the $k^{\textrm{th}}$ acoustic reflection observed by the $m$ and $n^{\textrm{th}}$ sensor, respectively. The general frequency weighting is indicated by $\Psi(\omega)$ \cite{knapp1976generalized}. Since acoustic reflections are assumed to exhibit broadband frequency spectra, the general frequency weighting can be set to $\Psi(\omega) \equiv 1$, reducing the \gls{gcc} to regular cross-correlation. In \citet{rosseel2021improved}, it was shown that the cross-correlation between two sinc functions, as defined for acoustic reflections in (\ref{eq:tdeinterp2024:sigmodel_bandlimit_acoust_refl}), is a scaled and time-delayed sinc function of the form

\begin{equation}
    \label{eq:tdeinterp2024:cross_corr_sinc}
    r^k_{m,n}(\tau) = \frac{\sinc\left(f_\mathrm{s} \left(t^k_m - t^k_n + \tau \right)\right)}{16 \pi^2 f_\mathrm{s} \, c^2 \, t^k_n \, t^k_m}.
\end{equation}

The \gls{tdoa} between two acoustic reflections measured at different points in space, can then be found as the time lag for which $r_{m,n}^k(\tau)$ is maximized, i.e.,

\begin{align}
    \label{eq:tdeinterp2024:max_gcc}
    \hat\tau_{m,n}^k & = \arg\max_{\tau} r^k_{m,n}(\tau) = t^k_n - t^k_m.
\end{align}

Moreover, when image-sources are assumed to be located in the far field relative to the sensor array position, the plane wave propagation model can be assumed for the localization of the image-sources. This assumption generally leads to a more efficient method for image-source localization, and makes it possible to rewrite the \gls{tdoa} information using the \gls{sv} $\mathbf{s}^k$ relating to an image-source with index $k$ \cite{pirinen2009confidence}, such that

\begin{equation}
    \tau_{m,n}^k = \mathbf{v}_{m,n}^T \mathbf{s}^k,
\end{equation}

\noindent where $\mathbf{v}_{m,n}$ is the sensor position difference vector, defined as

\begin{equation}
    \mathbf{v}_{m,n} = \mathbf{r}_n - \mathbf{r}_m.
\end{equation}

The \gls{sv} fully describes the propagation characteristics of a plane wave, and is defined as the vector that points in the direction of the acoustic reflection $k$ with a magnitude equal to the inverse of the speed of sound $c$ in the medium. By estimating the \gls{tdoa} information using (\ref{eq:tdeinterp2024:max_gcc}) for all sensor pairs in the array, the estimated \gls{sv} $\hat{\mathbf{s}}^k$ can be found from the least squares solution \cite{pirinen2009confidence} given by

\begin{equation}
    \label{eq:tdeinterp2024:slowness_estimate}
    \hat{\mathbf{s}}^k = \mathbf{V}^+ \boldsymbol{\hat\tau}^k
\end{equation}

\noindent where $(.)^+$ denotes the Moore-Penrose pseudo-inverse, $\mathbf{V}$ is the matrix containing the sensor position difference vectors $\mathbf{v}_{m,n}$ for all pairs across the $N$ sensors in the array,

\begin{equation}
    \mathbf{V} = \left[\mathbf{v}_{1, 2}, \mathbf{v}_{1, 3}, \dots, \mathbf{v}_{N-1, N} \right]^T,
\end{equation}

\noindent and $\boldsymbol{\hat\tau}^k$ is the corresponding vector of \gls{tdoa} estimates,

\begin{equation}
    \boldsymbol{\hat\tau}^k = \left[\hat\tau_{1, 2}^k, \hat\tau_{1, 3}^k, \dots, \hat\tau_{N-1, N}^k \right]^T.
\end{equation}

From the estimated \gls{sv}, the position of each image-source $k$ can be found relative to the position of the sensor array's center $\mathbf{r}_c$, as

\begin{align}
    \label{eq:tdeinterp2024:im_position_est}
    \mathbf{\hat x}_k = \mathbf{r}_c - \frac{\mathbf{\hat s}^k \|\mathbf{r}_c - \mathbf{x}_k\|_2}{\|\mathbf{\hat s}^k \|_2},
\end{align}

\noindent where $\mathbf{r}_c = \frac{1}{N} \sum\limits_{n = 1}^{N} \mathbf{r}_n$ \cite{pirinen2009confidence}. The distance $\|\mathbf{r}_c - \mathbf{x}_k\|_2$ between the sensor array center and the $k^{\textrm{th}}$ image-source can be found from the \gls{toa} information $t_c^k$ at the center of the sensor array. If no central sensor is present in the array, the center \gls{toa} information can be estimated by averaging the \gls{toa} information from all sensors in the array, i.e.,

\begin{equation}
    \label{eq:tdeinterp2024:toa_center}
    t_c^k = \frac{1}{N} \sum_{n = 1}^{N} t_n^k.
\end{equation}

\subsection{Localization limitations}
Various factors can impact the performance of image-source localization methods \cite{tervo2013sdm, tervo2012acoustic, amengual2021optimizations}, including the presence of measurement noise, the type of excitation signal used for the \gls{rir} measurement, the type of sensor-array, the sampling rate, and the size of the analysis window. In this section, we will focus on two of these factors, namely the analysis window size, and the sampling rate.

\subsubsection{Analysis window size}
\label{sec:tdeinterp2024:win_size}

As briefly mentioned in the introduction, for acoustic reflector localization in a reverberant environment, the analysis window size should be chosen greater than the time it takes for an acoustic reflection wavefront to propagate through the sensor array, i.e.,

\begin{equation}
    \label{eq:tdeinterp2024:min_winsize}
    L > \frac{d_{\mathrm{max}}}{c} f_\mathrm{s},
\end{equation}

\noindent where $L$ is the analysis window length in samples and $d_{\mathrm{max}}$ is the maximum pairwise distance across all sensors in the array. The minimum window length $L$ ensures that an acoustic reflection will be captured by all sensors in the array, enabling the estimation of \gls{tdoa} information between all pairs of sensors. In the literature, the minimum window length is often defined as $L > 2 \frac{d_{\mathrm{max}}}{c} f_\mathrm{s}$, to ensure that a sufficient number of samples is available at every sensor to perform acoustic reflector localization \cite{tervo2013sdm, puomio2021parametric}. Whereas (\ref{eq:tdeinterp2024:min_winsize}) provides a valid lower bound, an increase in the number of samples within the analysis window will result in an improvement in localization performance. However, this also increases the likelihood of two acoustic reflections occurring within a single window, which will produce inaccurate time delay estimates leading to poorer localization performance. The choice of window size is therefore a trade-off between localization accuracy and the number of reflections that can be accurately localized in a single \gls{rir} measurement \cite{tervo2013sdm}. A maximum window size can heuristically be derived based on the echo density of the room \cite{tervo2012acoustic,kuttruff2017room}.

\subsubsection{Sampling rate}

The accuracy of image-source localization, as described by (\ref{eq:tdeinterp2024:slowness_estimate}) and (\ref{eq:tdeinterp2024:im_position_est}), is dependent upon the accuracy of \gls{toa} and \gls{tdoa} estimation. The \gls{toa} of acoustic reflections is typically determined by applying a peak-picking algorithm to the \gls{rir} \cite{defrance2008detecting, defrance2009using,antonacci2012inference, tervo2012acoustic, remaggi2017acoustic, diego2021dechorate}, while the \gls{tdoa} estimates are obtained by identifying the lag corresponding to the maximum sample value of the cross-correlation function $r_{m,n}^k(\tau)$, as per (\ref{eq:tdeinterp2024:max_gcc}). The accuracy of these estimates is therefore directly related to the sampling period $T = \frac{1}{f_\mathrm{s}}$ that is used during the \gls{rir} measurements.

In both \gls{toa} and \gls{tdoa} estimation, the true time delay typically does not coincide with an integer multiple of the sampling period $T$. Instead, it lies within the time interval $[(\kappa_0 - 1)T, (\kappa_0 + 1)T]$, where $\kappa_0$ denotes the index of the sample value that corresponds to the closest integer multiple of the true time delay. Consequently, the estimated time delays are subject to a maximum discretization error of $T$ seconds.

To decrease the discretization error that is made during \gls{toa} and \gls{tdoa} estimation, the sampling rate $f_\mathrm{s}$ can be increased, resulting in a smaller sampling period $T$. However, hardware limitations during \gls{rir} measurements often restrict the achievable sampling rate. In such cases, interpolation schemes can be applied to enable \gls{toa} and \gls{tdoa} estimation with subsample accuracy. The following section will review several existing interpolation schemes for improving \gls{tde} with subsample accuracy. \section{Time-delay interpolation}
\label{sec:tdeinterp2024:interpolation}

The discretization error in \gls{tde} can be mitigated by applying time-interpolation methods to a discretized \gls{tde} function. For \gls{toa} estimation of acoustic reflections, this function corresponds to the filter-matched \gls{rir}, while for \gls{tdoa} estimation, it corresponds to the cross-correlation function between windowed \glspl{rir} measured at two spatially-separated sensors.  This section will discuss existing interpolation methods, including parabolic interpolation \cite{lai1999interpolation}, Gaussian interpolation \cite{zhang2005cross}, frequency interpolation \cite{svilainis2013subsample, svilainis2019review}, and sinc interpolation \cite{rosseel2021improved}, which was proposed by the authors. Finally, Whittaker-Shannon interpolation is discussed as a generalization of sinc interpolation for \gls{tde} in reverberant conditions.

\subsection{Parabolic Interpolation}

Parabolic interpolation is a three-point curve-fitting interpolation method used to improve \gls{tde} with subsample accuracy \cite{lai1999interpolation}. This method involves finding a quadratic function which passes through the maximum sample value of the \gls{tde} function and its two neighboring points. The quadratic function has the form $y_\mathrm{p}(\uptau) = a\uptau^2 + b\uptau + c$. The improved time delay estimate $\tilde\uptau_\mathrm{p}$ can be found at the vertex of the fitted parabolic curve. The symbol $\uptau$ is defined as a generalized time symbol $\uptau$, which represents time for \gls{toa} estimation, and time lag for \gls{tdoa} estimation. For a discretized function $f[\kappa]$, with a maximum sample value at the discrete index $\kappa_0$, the vertex position can be found at \cite{lai1999interpolation},

\begin{equation}
    \tilde\uptau_\mathrm{p} = \frac{f[\kappa_0 - 1] - f[\kappa_0 + 1]}{2 \left(f[\kappa_0 - 1] - f[\kappa_0] + f[\kappa_0 + 1]\right)}.
\end{equation}

Parabolic fitting is a popular interpolation method for subsample \gls{tde} because of its simplicity and computational efficiency. However, a major drawback is the high estimation bias it produces \cite{cespedes1995methods}. Estimation bias refers to the systematic error in the estimated time delay. The magnitude and direction of this bias depend on the exact position of the true time delay relative to the sampling period $T_\mathrm{s} = 1 / f_\mathrm{s}$, i.e., where the true time delay falls within $T_\mathrm{s}$ \cite{cespedes1995methods}.

\subsection{Gaussian Interpolation}

Alternatively, an improved time delay estimate can be found by fitting the \gls{tde} function with an exponential function of the form $y_\mathrm{g}(\uptau) = \alpha \cdot e^{-\beta(\uptau - \tilde\uptau_\mathrm{g})^2}$. The vertex position of this function depends only on parameter $\tilde\uptau_\mathrm{g}$, which can be found from the maximum sample value of the discretized function and its neighboring points \cite{zhang2005cross} as,

\begin{equation}
    \tilde\uptau_\mathrm{g} = \frac{\ln(f[\kappa_0 + 1]) - \ln(f[\kappa_0 - 1])}{4 \ln(f[\kappa_0]) - 2 \ln(f[\kappa_0 + 1]) + \ln(f[\kappa_0 - 1])},
\end{equation}

\noindent where it is required that $f[\kappa] > 0$, for $\kappa \in \{\kappa_0-1, \kappa_0, \kappa_0 + 1\}$. If this condition is not met, as can happen with critically sampled functions, \cite{rosseel2021improved} proposes to estimate $\tilde\uptau_\mathrm{g}$ by first offsetting the values $f[\kappa_0-1], f[\kappa_0], f[\kappa_0 + 1]$, by a common scalar constant to ensure that they are strictly positive. 

This interpolation approach is utilized in the \gls{sdm} \cite{tervo2013sdm, meyerkahlen2022what} to enhance the localization accuracy of acoustic reflectors and facilitate spatial sound field analysis. Compared to parabolic interpolation, Gaussian interpolation generally produces a lower \gls{tdoa} estimation error, but remains susceptible to estimation bias \cite{cespedes1995methods}.

\subsection{Frequency Interpolation}

While the previous interpolation methods are performed in the time-domain, and can be applied to both \gls{toa} and \gls{tdoa} estimation, frequency interpolation is specifically designed for improving \gls{tdoa} estimation in the frequency domain \cite{svilainis2013subsample, svilainis2019review}. Frequency interpolation relies on the observation that the phase spectrum of the \gls{csd} function, which is the frequency-domain transformation of the cross-correlation function obtained through a \gls{dtft}, is linear due to the complex conjugation operation of the \gls{csd}. The method works by first eliminating phase-wrapping by circularly shifting the cross-correlation function in the time-domain by a rough estimate $\kappa_0$ of the time delay \cite{svilainis2013subsample}.

\begin{equation}
    \hat{f}[\kappa] = \textrm{circshift}(f[\kappa], \kappa_0).
\end{equation}

The shifted version of the \gls{tde} function $\hat{f}[k]$ can now be transformed to the frequency domain to obtain the \gls{csd} $\hat{F}[\omega]$ for \gls{tdoa} estimation. The subsample \gls{tdoa} estimate $\tilde\uptau_\mathrm{f}$ corresponds to the slope of a linear regression fit on the phase values. It was shown in \cite{svilainis2019review}, that minimizing the L2 norm of the difference between the \gls{csd} and a linear function of the form $\alpha\omega$ exhibited the best performance. The improved \gls{tdoa} estimate using frequency interpolation can be found as,

\begin{equation}
    \tilde\uptau_\mathrm{f} = \kappa_0 + \min_\alpha \left\| |\hat{F}[\omega]|^2 \left(\varphi(\hat{F}[\omega]) - \alpha\omega\right) \right\|_2,
\end{equation}

\noindent where $\varphi(\cdot)$ denotes the phase operator and $\omega$ denotes frequency.

Frequency interpolation exhibits very low estimation bias \cite{svilainis2019review}. However, its accuracy is dependent upon sufficiently high \gls{snr} conditions. This dependency can become a limitation when localizing acoustic reflections from distant image-sources relative to the sensor array, as the \gls{snr} of such reflections diminishes according to the inverse square law of sound propagation. Additionally, since frequency interpolation depends on the linearity of the phase in the frequency-domain \gls{tde} function, it is unsuitable for \gls{toa} estimation with subsample precision. However, this method can provide reasonable accuracy for \gls{toa} estimation when the phase of the \gls{tde} function remains approximately linear. In \Cref{sec:tdeinterp2024:sim_results}, the performance of frequency interpolation is evaluated for \gls{tdoa} and \gls{toa} estimation of acoustic reflections in reverberant conditions.

\subsection{Sinc Interpolation}

In \cite{rosseel2021improved}, the authors proposed an alternative interpolation method for improving \gls{tde} of acoustic reflections. This method finds its basis in the Nyquist-Shannon sampling theory and takes advantage of the assumption that the \gls{tde} function for \gls{toa} and \gls{tdoa} estimation of a single acoustic reflection can be modeled as a sinc function. For \gls{toa} estimation, this is evident from (\ref{eq:tdeinterp2024:sigmodel_bandlimit_acoust_refl}). For \gls{tdoa} estimation, it was shown in (\ref{eq:tdeinterp2024:cross_corr_sinc}) that the cross-correlation function between two sinc functions is a scaled and time-delayed sinc function with a time delay equal to the \gls{tdoa}. From this observation, it can be inferred that sinc interpolation is the most suitable interpolation method for \gls{tdoa} estimation of acoustic reflections.

Sinc interpolation consists of fitting a critically sampled and truncated sinc function around the maximum sample value of the normalized \gls{tde} function. The fractional part of the time delay can be found by minimizing the following objective function in the discrete-time domain,

\begin{equation}
    \label{eq:tdeinterp2024:sinc_int_costfun}
    \kappa_\mathrm{s} = \arg \min_{\kappa_\mathrm{s}} \sum^{\kappa_0 + S}_{\kappa = \kappa_0 - S} \left| \sinc(\pi f_\mathrm{s} (\kappa T - \kappa_s T_\mathrm{i})) - \frac{f[\kappa]}{f[\kappa_0]} \right|^2,
\end{equation}

\noindent where $\kappa_\mathrm{s}$ is the fractional part of the time delay, $S$ is a tunable parameter determining the truncation point of the sinc function, and $T_\mathrm{i}$, where $T_\mathrm{i} < T$, represents the interpolation sampling period which determines the resolution of the interpolation. The time delay with subsample accuracy is found as $\tilde\uptau_\mathrm{s} = \kappa_0 T + \kappa_\mathrm{s} T_\mathrm{i}$.

It was shown in \cite{rosseel2021improved} that this method outperforms both parabolic and Gaussian interpolation for \gls{tdoa} estimation of acoustic reflections in the free-field. However, the performance of this method has not yet been evaluated for \gls{toa} estimation and acoustic reflector localization in reverberant conditions, where the measured \glspl{rir} are typically windowed to isolate each acoustic event. Moreover, sinc interpolation has not been compared to frequency interpolation for \gls{tdoa} and \gls{toa} estimation. In the following section, an adaptation of sinc interpolation is presented for acoustic reflector localization in reverberant conditions, where the acoustic events are isolated using an analysis window.
 \vspace{-1em}
\section{Applying a windowing function}
\label{sec:tdeinterp2024:windowing}

In the previous sections, it was assumed that each acoustic reflection $h^k_n(t)$ could be perfectly separated in time. In practice however, we only have access to the \gls{rir} $h_n(t)$, which is a superposition of all acoustic reflections arriving at the sensor array. Isolating individual acoustic reflections is not trivial and poses significant challenges.

To distinguish between individual acoustic reflections in the \gls{rir}, a sliding short-time windowing function can be applied to $h_n(t)$ \cite{tervo2012acoustic}. This analysis window, denoted by $\gamma(t, v)$, is only non-zero within the interval $t \in [v - \frac{L}{2} \dots v + \frac{L}{2}] T$, where $v$ is the window position and $L$ denotes the window length in samples. The window length is chosen based on the physical dimensions of the sensor array, see section \ref{sec:tdeinterp2024:win_size}. By applying the analysis window at every discrete time instant $vT$, with $v \in \mathbb{Z}$, a set of short-time windowed \glspl{rir} $\tilde{h}_n(t, v)$ is obtained which partially overlap in time. It is assumed that there is at most a single acoustic reflection $k$ present in each window \cite{tervo2013sdm}. This assumption generally holds for the direct path and early reflections of the \gls{rir} $h_n(t)$, provided the sensor positioning avoids geometric symmetries with respect to the sound source and surrounding room boundaries. The short-time windowed \glspl{rir} are given by,

\begin{equation}
    \tilde{h}_n(t, v) = h_n(t) \gamma(t, v)
\end{equation}

\noindent where $v$ will be referred to as the window number.

The \gls{toa} and \gls{tdoa} of the acoustic reflections can be estimated by analyzing the short-time windowed \glspl{rir}. The \gls{toa} can be estimated from each discretized short-time windowed \gls{rir} $\tilde{h}_n[\kappa, v]$ by finding the sample-index of the maximum absolute amplitude. The \gls{toa} $\hat{t}_n(v)$ of the acoustic reflection in window $v$ is estimated by,

\begin{equation}
    \hat{t}_n(v) = \frac{v + \arg\max_\kappa |\tilde{h}_n[\kappa, v]|}{f_\mathrm{s}}.
\end{equation}

Similarly, the \gls{tdoa} $\hat{\tau}_{m,n}(v)$ of an acoustic reflection in window $v$ is estimated as the time-index which maximizes the cross-correlation between two short-time windowed \glspl{rir} $\tilde{h}_n(t, v)$ and $\tilde{h}_m(t, v)$, where $m$ denotes the reference sensor. The \gls{tdoa} $\hat{\tau}_{m,n}(v)$ is estimated by,

\begin{equation}
    \hat{\tau}_{m,n}(v) = \frac{\arg\max_\kappa r_{m,n}[\kappa, v]}{f_\mathrm{s}},
\end{equation}

\noindent where $r_{m,n}[\kappa, v]$ denotes the cross-correlation between the discretized short-time windowed \glspl{rir} $\tilde{h}_m[\kappa, v]$ and $\tilde{h}_n[\kappa, v]$. \section{Reformulating sinc interpolation}
\label{sec:tdeinterp2024:sinc_reformulation}

By applying an analysis window to the measured \glspl{rir}, the cross-correlation between two windowed acoustic reflections deviates from the scaled sinc function previously described in (\ref{eq:tdeinterp2024:cross_corr_sinc}). This deviation motivates the need for a reformulation of sinc interpolation to accommodate \gls{tde} of windowed acoustic reflections with subsample accuracy. The performance implications of this windowing operation on the previously proposed sinc interpolation are examined in \Cref{sec:tdeinterp2024:sim_results}. This section presents a reformulation of sinc interpolation for short-time windowed \gls{toa} and \gls{tdoa} estimation, resulting in a method known as Whittaker-Shannon interpolation which has not been previously applied to \gls{tde} of acoustic reflections.

It can be shown that Whittaker-Shannon interpolation can be derived from the previously proposed sinc interpolation. The objective function in (\ref{eq:tdeinterp2024:sinc_int_costfun}) can be adapted by using the windowed \gls{tde} function $f(\kappa T, v)$, which is the cross-correlation function of the windowed \glspl{rir} for \gls{tdoa} estimation between two sensors, and the windowed \gls{rir} for \gls{toa} estimation. The resulting objective function is given by

\begin{equation}
    \kappa_\mathrm{ws} = \arg \min_{\kappa_\mathrm{ws}} \sum_{\kappa =-\infty}^{\infty} \left| \sinc(\pi f_\mathrm{s} (\kappa T - \kappa_\mathrm{ws} T_\mathrm{i})) - f(\kappa T, v) \right|^2.
\end{equation}

No truncation is applied to the objective function and $f(\kappa T, v)$ is not normalized by the maximum value of the \gls{tde} function, as in (\ref{eq:tdeinterp2024:sinc_int_costfun}). The objective function can be reduced by completing the square, as

\begin{align}
    \kappa_\mathrm{ws} = &\arg\min_{\kappa_\mathrm{ws}} \Big( \sum_{\kappa=-\infty}^\infty \sinc\left(\pi f_\mathrm{s} \left(\kappa T - \kappa_\mathrm{ws}T_\mathrm{i}\right)\right)^2 \notag \\
    & -2 \sum_{\kappa=-\infty}^\infty f(\kappa T, v) \sinc\left(\pi f_\mathrm{s} \left(\kappa T - \kappa_\mathrm{ws}T_\mathrm{i}\right)\right) \notag \\
    & + f(\kappa T, v)^2\Big),
\end{align}

\noindent where it can be shown that the first term is equal to 1, and the third term is independent of $\kappa_\mathrm{ws}$, leading to the following reduced objective function:

\begin{align}
    \kappa_\mathrm{ws} & = \arg\min_{\kappa_\mathrm{ws}} \sum_{\kappa=-\infty}^{\infty} - f(\kappa T, v) \, \sinc(\pi f_\mathrm{s} (\kappa T - \kappa_\mathrm{ws} T_\mathrm{i}))  \notag \\
                       & = \arg\max_{\kappa_\mathrm{ws}} \sum_{\kappa=-\infty}^{\infty} f(\kappa T, v) \, \sinc(\pi f_\mathrm{s} (\kappa_\mathrm{ws} T_\mathrm{i} - \kappa T)),
\end{align}

\noindent where $f(\kappa T, v)$ is only non-zero in the interval $\kappa \in \mathcal{L}$. The objective function can be further reduced by changing the bounds of the summation to

\begin{equation}
    \kappa_\mathrm{ws} = \arg\max_{\kappa_\mathrm{ws}} \sum_{\kappa \in \mathcal{L}} f(\kappa T, v) \, \sinc(\pi f_\mathrm{s} (\kappa_\mathrm{ws} T_\mathrm{i} - \kappa T)),
\end{equation}

\noindent where $\mathcal{L} = [-\frac{L}{2}, \frac{L}{2}]$ for \gls{toa} estimation, and $\mathcal{L} = [-L + 1, L - 1]$ for \gls{tdoa} estimation. 

This result shows that sinc interpolation can be reformulated to a discrete convolution of the \gls{tde} function $f(\kappa T, v)$ and a critically sampled sinc function by assuming infinite support of the sinc function over discrete time. This finding can be related back to Whittaker-Shannon interpolation \cite{marks1991introduction}, which is derived from the Nyquist-Shannon sampling theorem \cite{shannon1949communication}. The Whittaker-Shannon interpolation method relies on the principle that a band-limited signal can be reconstructed from its discrete samples if the sampling frequency $f_\mathrm{s}$ is at least twice the highest frequency present in the signal \cite{shannon1949communication}. In this case, the sinc function acts as an interpolation kernel that reconstructs the continuous-time signal from its discrete samples.

The Whittaker-Shannon interpolation can be modified to only consider $S$ samples around the maximum value of the \gls{tde} function, as done in (\ref{eq:tdeinterp2024:sinc_int_costfun}). The resulting optimization problem is given by

\begin{equation}
    \label{eq:tdeinterp2024:sinc_int_costfun_win}
    \kappa_\mathrm{ws} = \arg\max_{\kappa_\mathrm{ws}} \sum_{\kappa=\kappa_0 - S}^{\kappa_0 + S} f(\kappa T, v) \, \sinc(\pi f_\mathrm{s} (\kappa_\mathrm{ws} T_\mathrm{i} - \kappa T)).
\end{equation}

\begin{figure}[ht]
    \centering
    \includegraphics[width=\linewidth]{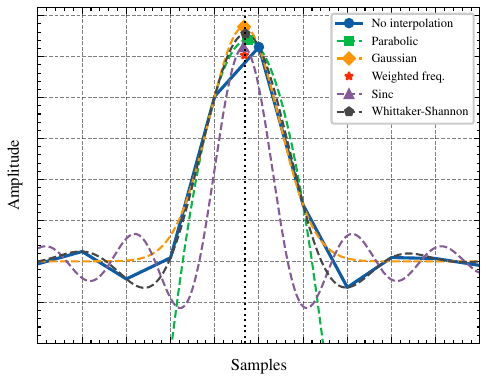}
    \caption{Time-domain visualization of interpolation methods applied to a simulated acoustic reflection, modeled using a Thiran fractional delay filter with a band-limit of $B = \frac{2 f_\mathrm{s}}{5}$. The dotted vertical line represents the true time delay, and the marker shows the estimated maximum value obtained using each interpolation method. For weighted frequency-domain interpolation, only the peak value is included, as the method does not operate on the time-domain signal.}
    \label{fig:tdeinterp2024:interp_bandlimited}
\end{figure}

The interpolation methods presented in this section are illustrated in the time domain in \Cref{fig:tdeinterp2024:interp_bandlimited}. In this figure, an acoustic reflection is simulated using a Thiran fractional delay filter, which is described in greater detail in \Cref{sec:tdeinterp2024:simulations}. A band-limit of $B = \frac{2 f_\mathrm{s}}{5}$ is imposed on the simulated reflection, after which various interpolation techniques are applied. For weighted frequency-domain interpolation, only the estimated peak value is shown, as this method does not operate directly in the time domain. The dotted vertical line marks the true time delay, while the marker indicates the location of the maximum value estimated by each interpolation method.

In the following section, the performance of the different interpolation methods is evaluated using simulated data. The results are presented in \Cref{sec:tdeinterp2024:sim_results}. \section{Simulations}
\label{sec:tdeinterp2024:simulations}

To objectively assess the performance of the various interpolation methods for acoustic reflection localization through subsample \gls{toa} and \gls{tdoa} estimation, a series of simulations were developed in Python \cite{rosseel2023improving_github}.

\subsection{Setup}

The simulations were designed to closely mimic real-world conditions while retaining the ability to objectively measure subsample \gls{toa} and \gls{tdoa} performance using different interpolation methods. For simplicity, the simulations were carried out in two dimensions only. This limitation in dimensionality does not impact the performance of the interpolation methods for \gls{toa} and \gls{tdoa} estimation.

The simulation setup consisted of a six-channel circular sensor array with a radius of 50 $\si{\milli\metre}$, positioned at the origin of the two-dimensional simulated space. The propagation speed of sound was set to $c = 343$ $\si{\metre\per\second}$, and the default sampling rate was set to $f_\mathrm{s} = 8$ $\si{\kilo\hertz}$. The signals simulated at each sensor position were corrupted with white noise to simulate measurement noise and had an \gls{snr} of 40 $\si{\deci\bel}$. To ensure that the \gls{snr} remained constant for all acoustic reflections in the simulation, the reflections were not scaled according to the distance between each sensor and the simulated image source. The analysis window had a duration of 4 $\si{\milli\second}$, which at the default sampling rate results in a window length of $L = 32$ samples. The time-resolution with which sinc and Whittaker-Shannon interpolation methods were applied was determined by the interpolation factor $i = T / T_\mathrm{i}$, where $T$ is the operating sampling period and $T_\mathrm{i}$ is the interpolation sampling period used in sinc and Whittaker-Shannon interpolation. The default interpolation factor was set to $i = 200$ during the simulations.

It was ensured that, for each window of length $L$, only a single acoustic wavefront originating from an image source was present. This was accomplished by positioning $K$ image sources around a sensor array in two-dimensional space such that the spherical wave originating from each image source arrived at the center of the sensor array at different time instants with a time lag of $L / f_\mathrm{s}$ seconds between each arrival. The position of the image sources can be described in two-dimensional space by the following equation:

\begin{equation}
    \mathbf{x}_k = \mathbf{r}_c +
    \begin{bmatrix}
        \cos(\theta_k) \\
        \sin(\theta_k)
    \end{bmatrix} \cdot k\frac{c L}{f_\mathrm{s}},
\end{equation}

\begin{figure}[ht]
    \resizebox{\columnwidth}{!}{\includegraphics{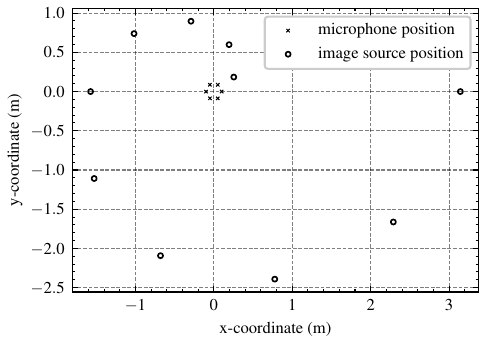}
    }
    \caption{Positioning of ten image sources around a sensor array such that the time of arrival in the center of the sensor array is equal to $t_k = k \cdot L / f_\mathrm{s}$. For illustration purposes, the angular direction $\theta_k$ of each image source was set to $\theta_k = k \cdot 2 \pi / K$.}
    \label{fig:tdeinterp2024:sim_setup}
\end{figure}

\noindent where $\theta_k$ is the angle of the $k$th image source relative to the center of the sensor array. This is shown in \Cref{fig:tdeinterp2024:sim_setup}, where ten image sources are placed around a sensor array with a \gls{toa} at the center of the sensor array equal to $t_k = k \cdot L / f_\mathrm{s}$. In the simulations, the angular direction $\theta_k$ of each image source relative to the center of the sensor array was randomly distributed between $\theta_k \in [0, 2\pi]$ according to a uniform distribution. The number of image sources was set to $K = 1000$.

A number of independent parameters were varied to assess their influence on the performance of the interpolation methods for \gls{tde} with subsample accuracy. The independent parameters included the sampling frequency, the interpolation factor $i$, the \gls{snr}, and the analysis window length. When varying each parameter, all other parameters were kept constant. The performance of the interpolation methods was evaluated in terms of the \gls{toa} and \gls{tdoa} error, which was calculated as the absolute difference between the estimated and true \gls{toa} and \gls{tdoa}, respectively. The \gls{tdoa} error was averaged over all sensor pairs and image-sources, while the \gls{toa} error was averaged over all sensors and image-sources.

The sampling frequency was varied from 2 $\si{\kilo\hertz}$ to 48 $\si{\kilo\hertz}$, which allowed for an analysis into the effects of different sampling rates on the accuracy of \gls{tde} using different interpolation factors. The interpolation factor $i$ was varied from $1$ to $200$ to analyse the performance of the sinc and Whittaker-Shannon interpolation methods under different time-resolution conditions.  The robustness of each interpolation method under varying noise conditions was investigated by varying the \gls{snr} from $-10$ $\si{\deci\bel}$ up to $60$ $\si{\deci\bel}$. Finally, to determine the influence of the analysis window on \gls{tdoa} estimation, the window duration was varied from $1$ $\si{\milli\second}$ up to $16$ $\si{\milli\second}$, which at a sampling rate of $f_\mathrm{s} = 8$ $\si{\kilo\hertz}$ corresponds to a window length of $L = 8$ samples up to $L = 128$ samples.

For each simulation, the time delays were calculated at the time instants when an acoustic reflection arrived at the middle of the sensor array. This ensures that the time delay estimates were calculated at the time instant when each windowed sensor signal contains at most a single acoustic reflection. The simulation results are discussed in section \ref{sec:tdeinterp2024:sim_results}.

\subsection{Modeling Fractional Delays}

The acoustic waves emanating from each image source towards the sensors were modeled using spherical wave propagation. As the \gls{toa} of each acoustic wavefront is not necessarily an integer multiple of the sampling period, a fractional delay filter was employed to simulate the true \gls{toa} of each acoustic wave with subsample accuracy. Ideally, a shifted sinc function serves as the optimal filter for modeling fractional delays \cite{laakso1996splitting}. However, its use in the simulations would introduce an advantage to sinc and Whittaker-Shannon interpolation methods. To ensure a fair evaluation across various interpolation techniques, the Thiran all-pass filter \cite{laakso1996splitting} was selected. Thiran all-pass filters are a type of IIR fractional delay filter that exhibit a maximally flat group delay response at DC. The transfer function of a $P^{\textrm{th}}$ order Thiran filter is given by

\begin{align}
    H(z) & = \frac{z^{-P} A(z^{-1})}{A(z)} \notag                                                                                      \\
         & = \frac{a_P + a_{P-1} z^{-1} + \dots + a_1 z^{-(P - 1)} + z^{-P}}{1 + a_1 z^{-1} + \dots + a_{P-1}z^{-(P-1)} + a_P z^{-P}},
\end{align}

\noindent where for a desired fractional delay of $\delta$ samples, resulting in a total filter delay $\Delta = P + \delta$ samples, the coefficients $a_p$ for $p = 0,1,2,\dots,P$ are given by

\begin{equation}
    a_p = (-1)^p \binom{P}{p} \prod_{k=0}^P \frac{\delta-P + k}{\delta - P + p + k}.
\end{equation}

The precise modeling of fractional delays is crucial for evaluating the performance of the various interpolation methods. As demonstrated in \cite{valimaki1995discretetime}, the fractional delay modeling error of the Thiran filter is minimized for $P - \frac{1}{2} \leq \delta < P + \frac{1}{2}$. In order to ensure a minimal modeling error for arbitrary fractional delays $\delta$, the Thiran fractional delay filter is implemented in such a way that the desired fractional delay $\delta$ is independent of the filter order $P$, while also minimizing the modeled error. This is accomplished by decomposing the total filter delay $\Delta$ into an integer delay $\Delta_\mathrm{i}$ and a fractional delay $\Delta_\mathrm{f}$, as follows:

\begin{align}
    \Delta_\mathrm{i} & = \left\lfloor \Delta + \frac{1}{2} \right\rfloor, \\
    \Delta_\mathrm{f} & = \Delta - \Delta_\mathrm{i}.
\end{align}

When $\Delta_\mathrm{i} \ge P$, the total filter delay $\Delta$ can be implemented using a Thiran all-pass filter of order $P$. This is achieved by first applying an integer delay of $\Delta_\mathrm{i} - P$ to the signal, followed by the Thiran all-pass filter with a desired fractional delay of $\delta = \Delta_\mathrm{f}$ samples. This ensures that the output is delayed by exactly $\Delta$ samples. In the case where $\Delta_\mathrm{i} < P$, the total filter delay $\Delta$ can be implemented by first applying a fractional delay of $\delta = \Delta_\mathrm{f}$ samples using a $P^{\textrm{th}}$ order Thiran all-pass filter, after which the first $P - \Delta_\mathrm{i}$ samples of the output are removed.

\subsection{Design Parameter $S$}

\begin{figure}[ht]
        \resizebox{\columnwidth}{!}{\includegraphics{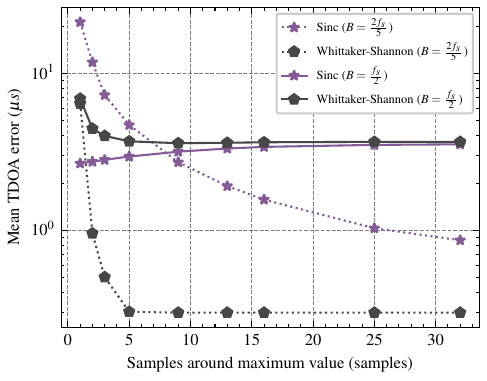}
        }
        \caption{The effect of varying parameter $S$ from $1$ sample up to $L$ samples around the maximum value on the performance of sinc and Whittaker-Shannon interpolation for \gls{tdoa} interpolation. This effect is examined under two bandwidth conditions: a critically sampled condition, $B = \frac{f_\mathrm{s}}{2}$, and one where the bandwidth is limited to $B = \frac{2 f_\mathrm{s}}{5}$. The non-varying parameters were set to $f_\mathrm{s} = 8$ $\si{\kilo\hertz}$, $i = 200$, \gls{snr} = 40 $\si{\deci\bel}$, and $L = 32$ samples.}
    \label{fig:tdeinterp2024:design_param_s_tdoa}
\end{figure}

\begin{figure}[ht]
    \begin{center}
        \resizebox{\columnwidth}{!}{\includegraphics*{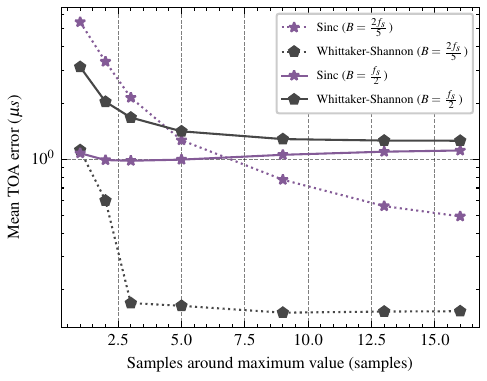}
        }
        \caption{The effect of varying parameter $S$ from $1$ sample up to $\frac{L}{2}$ samples around the maximum value on the performance of sinc and Whittaker-Shannon interpolation for \gls{toa} interpolation. This effect is examined under two bandwidth conditions: a critically sampled condition, i.e., $B = \frac{f_\mathrm{s}}{2}$, and one where the bandwidth is limited to $B = \frac{2 f_\mathrm{s}}{5}$. The non-varying parameters were set to $f_\mathrm{s} = 8$ $\si{\kilo\hertz}$, $i = 200$, \gls{snr} = 40 $\si{\deci\bel}$, and $L = 32$ samples.}
        \label{fig:tdeinterp2024:design_param_s_toa}
    \end{center}
\end{figure}

The design parameter $S$ specifies how many samples are included in sinc interpolation (\ref{eq:tdeinterp2024:sinc_int_costfun}) and Whittaker-Shannon interpolation (\ref{eq:tdeinterp2024:sinc_int_costfun_win}). The effect of this design parameter on the performance of the interpolation methods was assessed under two different bandwidth conditions: a critically sampled condition, i.e., $B = \frac{f_\mathrm{s}}{2}$, and one where the bandwidth is limited to $B = \frac{2 f_\mathrm{s}}{5}$. For \gls{tdoa} estimation, the design parameter $S$ was varied from $1$ sample up to $L$ samples for each bandwidth condition. For \gls{toa} estimation, the design parameter $S$ was varied from $1$ sample up to $\frac{L}{2}$ samples for each bandwidth condition. The results of this analysis are shown for \gls{tdoa} estimation in \Cref{fig:tdeinterp2024:design_param_s_tdoa}, and for \gls{toa} estimation in \Cref{fig:tdeinterp2024:design_param_s_toa}.

\subsubsection*{Critically Sampled Conditions}
In critically sampled conditions, where the bandwidth of the acoustic impulses arriving at the sensors is equal to the Nyquist frequency, i.e., $B = \frac{f_\mathrm{s}}{2}$, the performance of sinc interpolation is optimal for small values of $S$. This is because, for both \gls{toa} and \gls{tdoa} interpolation of critically sampled acoustic impulses, the \gls{tde} function is again a critically sampled sinc function. As a result, the best performance can be achieved when the interpolation is performed only on samples that are close to the maximum value of the estimation function. On the other hand, Whittaker-Shannon interpolation achieves optimal performance for \gls{tdoa} and \gls{toa} interpolation when parameter $S$ is set to a higher value. This is due to the fact that a critically sampled sinc function does not have sufficient samples available around the main lobe of the function to achieve an accurate interpolation using Whittaker-Shannon interpolation. Therefore, the best performance is achieved when additional samples around the maximum value are included in the interpolation.

\subsubsection*{Bandwidth Limited Conditions}
In band-limited conditions, where the bandwidth of the acoustic impulses arriving at the sensors is $B = \frac{2 f_\mathrm{s}}{5}$, sinc interpolation achieves optimal performance for \gls{tdoa} and \gls{toa} estimation when all samples in the window are included in the interpolation. In contrast, Whittaker-Shannon interpolation achieves optimal performance for both \gls{tdoa} and \gls{toa} estimation when $S$ is set to around half of the window size. Notably, the performance of both interpolation methods significantly deteriorates when $S$ is set to a small value.

The increase in performance of Whittaker-Shannon interpolation over sinc interpolation for \gls{toa} and \gls{tdoa} estimation in band-limited conditions can be attributed to the oversampling of the \gls{tde} function of band-limited acoustic reflections. This oversampling provides Whittaker-Shannon interpolation with a greater number of samples around the main lobe, allowing it to achieve more accurate \gls{toa} and \gls{tdoa} estimates.

A significant degradation in the performance of \gls{toa} and \gls{tdoa} estimation is observed in critically sampled conditions compared to band-limited conditions. This deterioration results from a limited availability of samples around the main lobe of a critically sampled sinc function, which negatively impacts the accuracy of \gls{toa} and \gls{tdoa} estimates. As a result, the performance of interpolation methods degrades in critically sampled conditions.

\subsection{Simulation Results}
\label{sec:tdeinterp2024:sim_results}

\begin{figure*}[t]
    \centering
    \subfigure[]{
        \includegraphics[width=\subfigwidth]{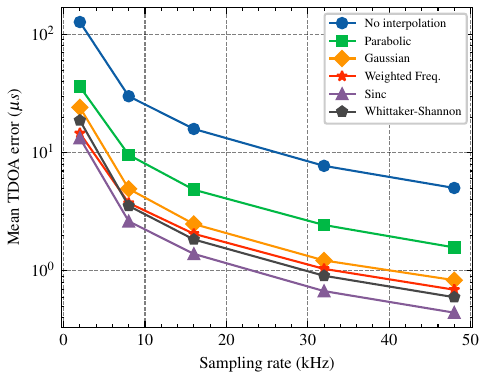}
        \label{fig:tdeinterp2024:sim_fs_tdoa_crit}}
    \subfigure[]{
        \includegraphics[width=\subfigwidth]{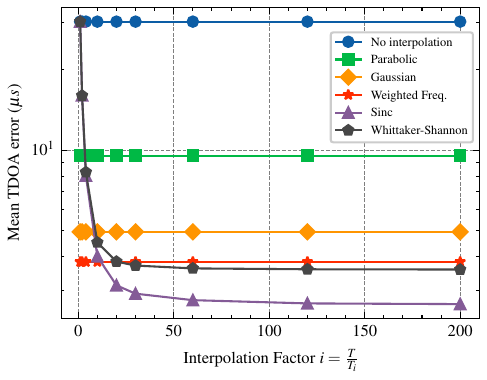}
        \label{fig:tdeinterp2024:sim_interp_tdoa_crit}} \\
    \subfigure[]{
        \includegraphics[width=\subfigwidth]{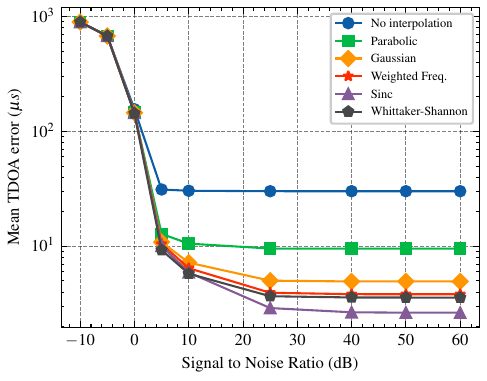}
        \label{fig:tdeinterp2024:sim_snr_tdoa_crit}}
    \subfigure[]{
        \includegraphics[width=\subfigwidth]{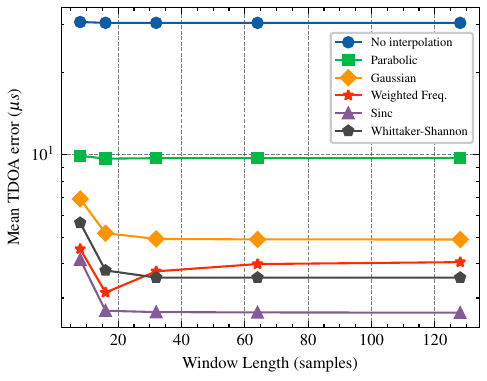}
        \label{fig:tdeinterp2024:sim_frame_len_tdoa_crit}}
    \caption{The effect of varying (a) the sampling frequency $f_\mathrm{s}$, (b) the interpolation factor $i$, (c) the \gls{snr}, and (d) the analysis window length $L$ on interpolation performance for \gls{tdoa} estimation of critically sampled acoustic reflections. The non-varying parameters were set to $f_\mathrm{s} = 8$ $\si{\kilo\hertz}$, $i = 200$, \gls{snr} = 40 $\si{\deci\bel}$, $L = 32$ samples, $S_\textrm{sinc} = 1$ sample, and $S_\mathrm{ws} = 9$ samples.}
    \label{fig:tdeinterp2024:sim_tdoa_crit}
\end{figure*}

\begin{figure*}[t]
    \centering
\subfigure[]{
        \includegraphics[width=\subfigwidth]{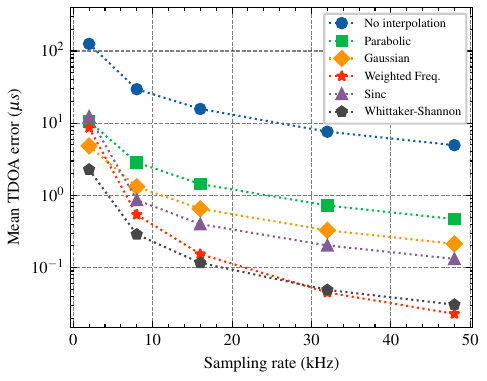}
        \label{fig:tdeinterp2024:sim_fs_tdoa_band}}
    \subfigure[]{
        \includegraphics[width=\subfigwidth]{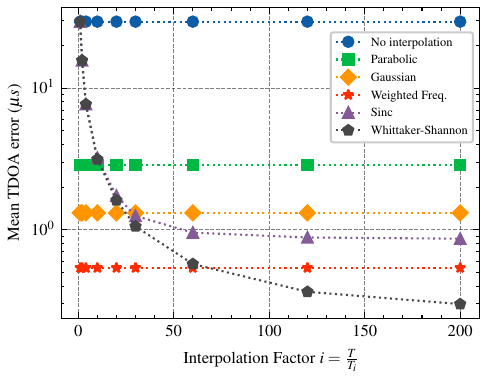}
        \label{fig:tdeinterp2024:sim_interp_tdoa_band}} \\
    \subfigure[]{
        \includegraphics[width=\subfigwidth]{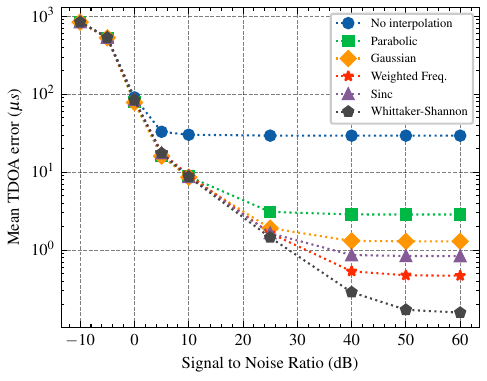}
        \label{fig:tdeinterp2024:sim_snr_tdoa_band}}
    \subfigure[]{
        \includegraphics[width=\subfigwidth]{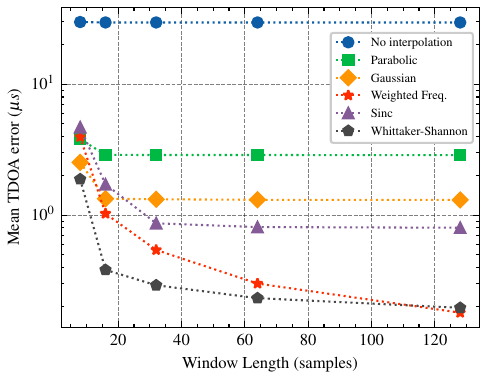}
        \label{fig:tdeinterp2024:sim_frame_len_tdoa_band}}
    \caption{The effect of varying (a) the sampling frequency $f_\mathrm{s}$, (b) the interpolation factor $i$, (c) the \gls{snr}, and (d) the analysis window length $L$ on interpolation performance for \gls{tdoa} estimation of band-limited acoustic reflections. The bandlimit was set to $B = \frac{2 f_\mathrm{s}}{5}$. The non-varying parameters were set to $f_\mathrm{s} = 8$ $\si{\kilo\hertz}$, $i = 200$, \gls{snr} = 40 $\si{\deci\bel}$, $L = 32$ samples, $S_\textrm{sinc} = 32$ samples, and $S_\mathrm{ws} = 9$ samples.}
    \label{fig:tdeinterp2024:sim_tdoa_band}
\end{figure*}

\begin{figure*}[ht]
    \centering
\subfigure[]{
        \includegraphics[width=\subfigwidth]{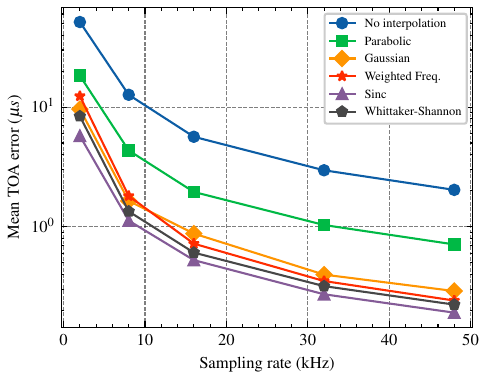}
        \label{fig:tdeinterp2024:sim_fs_toa_crit}}
    \subfigure[]{
        \includegraphics[width=\subfigwidth]{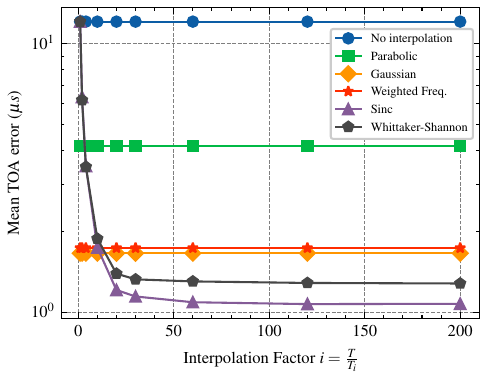}
        \label{fig:tdeinterp2024:sim_interp_toa_crit}} \\
    \subfigure[]{
        \includegraphics[width=\subfigwidth]{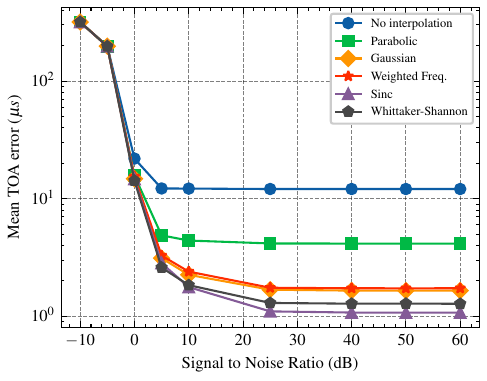}
        \label{fig:tdeinterp2024:sim_snr_toa_crit}}
    \subfigure[]{
        \includegraphics[width=\subfigwidth]{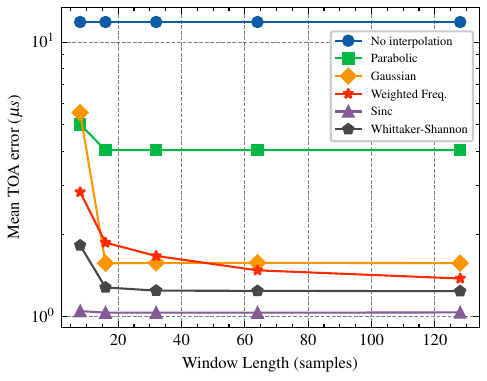}
        \label{fig:tdeinterp2024:sim_frame_len_toa_crit}}
    \caption{The effect of varying (a) the sampling frequency $f_\mathrm{s}$, (b) the interpolation factor $i$, (c) the \gls{snr}, and (d) the analysis window length $L$ on interpolation performance for \gls{toa} estimation of critically sampled acoustic reflections. The non-varying parameters were set to $f_\mathrm{s} = 8$ $\si{\kilo\hertz}$, $i = 200$, \gls{snr} = 40 $\si{\deci\bel}$, $L = 32$ samples, $S_\textrm{sinc} = 1$ sample, and $S_\mathrm{ws} = 9$ samples.}
    \label{fig:tdeinterp2024:sim_toa_crit}
\end{figure*}

\begin{figure*}[ht]
    \centering
\subfigure[]{
        \includegraphics[width=\subfigwidth]{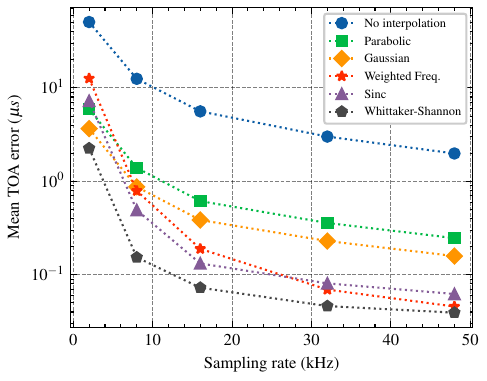}
        \label{fig:tdeinterp2024:sim_fs_toa_band}}
    \subfigure[]{
        \includegraphics[width=\subfigwidth]{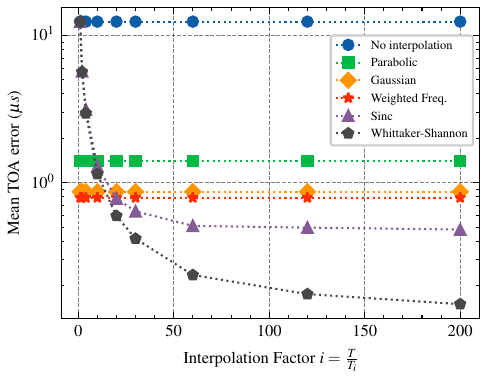}
        \label{fig:tdeinterp2024:sim_interp_toa_band}} \\
    \subfigure[]{
        \includegraphics[width=\subfigwidth]{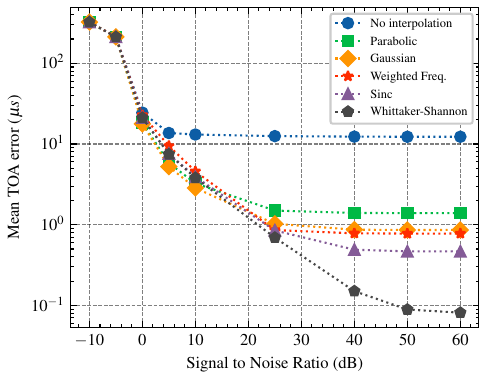}
        \label{fig:tdeinterp2024:sim_snr_toa_band}}
    \subfigure[]{
        \includegraphics[width=\subfigwidth]{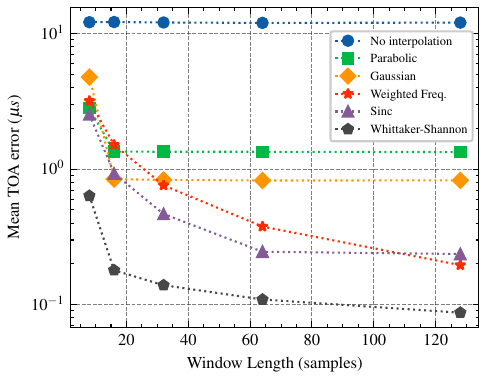}
        \label{fig:tdeinterp2024:sim_frame_len_toa_band}}
    \caption{The effect of varying (a) the sampling frequency $f_\mathrm{s}$, (b) the interpolation factor $i$, (c) the \gls{snr}, and (d) the analysis window length $L$ on interpolation performance for \gls{toa} estimation of band-limited acoustic reflections. The bandlimit was set to $B = \frac{2 f_\mathrm{s}}{5}$. The non-varying parameters were set to $f_\mathrm{s} = 8$ $\si{\kilo\hertz}$, $i = 200$, \gls{snr} = 40 $\si{\deci\bel}$, $L = 32$ samples, $S_\textrm{sinc} = 16$ samples, and $S_\mathrm{ws} = 9$ samples.}
    \label{fig:tdeinterp2024:sim_toa_band}
\end{figure*}

This section presents the simulation results, which demonstrate the impact of varying parameters on the performance of interpolation methods for \gls{toa} and \gls{tdoa} estimation. The effects of sampling rate, interpolation factor, \gls{snr}, and analysis window are discussed, with results illustrated in \Cref{fig:tdeinterp2024:sim_tdoa_crit} and \Cref{fig:tdeinterp2024:sim_tdoa_band} for \gls{tdoa} estimation with critically sampled and band-limited acoustic reflections, respectively. The results for \gls{toa} estimation are presented in \Cref{fig:tdeinterp2024:sim_toa_crit} and \Cref{fig:tdeinterp2024:sim_toa_band} for critically sampled and band-limited reflections, respectively.

\subsubsection{Sampling Rate}

The influence of sampling rate on \gls{tdoa} estimation performance is shown in \Cref{fig:tdeinterp2024:sim_fs_tdoa_crit} for critically sampled acoustic reflections and in \Cref{fig:tdeinterp2024:sim_fs_tdoa_band} for band-limited reflections. The results show that \gls{tdoa} estimation accuracy increases as sampling rates increase across all interpolation methods. This enhancement can be attributed to the higher time resolution achieved as the sampling period decreases.

For critically sampled reflections, sinc interpolation outperforms other methods. In contrast, Whittaker-Shannon interpolation excels for band-limited reflections at lower sampling frequencies. Notably, at very high sampling frequencies, weighted frequency interpolation becomes the top performer for band-limited reflections. This is because weighted frequency interpolation can effectively utilize the increased number of samples available within a fixed window duration at higher sampling rates, as observed in \Cref{fig:tdeinterp2024:sim_frame_len_tdoa_band}.

For \gls{toa} estimation, the performance of the interpolation methods is consistent with the results for \gls{tdoa} estimation. Sinc interpolation outperforms other methods for critically sampled reflections, while Whittaker-Shannon interpolation excels for band-limited reflections.

\subsubsection{Interpolation factor}

The impact of varying the interpolation factor on \gls{tdoa} estimation is shown in \Cref{fig:tdeinterp2024:sim_interp_tdoa_crit} and \Cref{fig:tdeinterp2024:sim_interp_tdoa_band} for critically sampled and band-limited acoustic reflections, respectively. It is shown that for critically sampled acoustic reflections, interpolation factors of $i \ge 20$ allow both sinc and Whittaker-Shannon interpolation to outperform other interpolation methods. Sinc interpolation achieves optimal performance for critically sampled reflections.

For band-limited reflections, it can be observed that Whittaker-Shannon interpolation outperforms weighted frequency interpolation from interpolation factors of $i \ge 60$. In both cases, as the interpolation factor increases, the performance of the interpolation methods increases. This is consistent with previous findings \cite{rosseel2021improved}.

For \gls{toa} estimation, the performance of the interpolation methods is consistent with the results for \gls{tdoa} estimation. Sinc interpolation outperforms other methods for critically sampled reflections, while Whittaker-Shannon interpolation excels for band-limited reflections.

\subsubsection{Signal-to-Noise Ratio}

Varying \gls{snr} significantly affects the performance of the interpolation methods for both \gls{toa} and \gls{tdoa} estimation. The influence of \gls{snr} on \gls{tdoa} estimation performance is demonstrated in \Cref{fig:tdeinterp2024:sim_snr_tdoa_crit} and \Cref{fig:tdeinterp2024:sim_snr_tdoa_band} for critically sampled and band-limited acoustic reflections, respectively. At low \gls{snr}, \gls{tdoa} estimation performance suffers significantly across all methods due to the noise-induced degradation of the cross-correlation function. In contrast, at higher \gls{snr} values, the performance trends observed earlier are reaffirmed. Specifically, sinc interpolation maintains its superiority over Whittaker-Shannon interpolation for critically sampled reflections, whereas Whittaker-Shannon interpolation achieves better performance for band-limited reflections.

The performance of \gls{toa} estimation when varying \gls{snr} is shown in \Cref{fig:tdeinterp2024:sim_snr_toa_crit} and \Cref{fig:tdeinterp2024:sim_snr_toa_band} for critically sampled and band-limited reflections, respectively. The results are similar to those observed for \gls{tdoa} estimation. When \gls{snr} is sufficiently high, sinc interpolation outperforms other methods for critically sampled reflections, while Whittaker-Shannon interpolation achieves better performance for band-limited reflections. At low \gls{snr}, the performance of all interpolation methods deteriorates significantly.

\subsubsection{Analysis Window Length}

The performance of \gls{tdoa} estimation when varying the analysis window length is shown in \Cref{fig:tdeinterp2024:sim_frame_len_tdoa_crit} and \Cref{fig:tdeinterp2024:sim_frame_len_tdoa_band} for critically sampled and band-limited acoustic reflections, respectively. It can be observed that for very small window lengths, the performance of \gls{tdoa} estimation degrades slightly for all interpolation methods due to insufficient information in the window. However, as the window length increases, the performance of \gls{tdoa} estimation converges for all interpolation methods. Notably, weighted frequency interpolation has an optimal window length for critically sampled reflections, but its performance deteriorates as the window length grows. In contrast, for band-limited reflections, the performance of weighted frequency interpolation increases with increasing window length, allowing it to outperform Whittaker-Shannon interpolation for very  long window lengths.

The performance of \gls{toa} estimation when varying the window length is shown in \Cref{fig:tdeinterp2024:sim_frame_len_toa_crit} and \Cref{fig:tdeinterp2024:sim_frame_len_toa_band} for critically sampled and band-limited reflections, respectively. The results are consistent with those observed for \gls{tdoa} estimation. For critically sampled reflections, sinc interpolation outperforms other methods, while Whittaker-Shannon interpolation excels for band-limited reflections.

\subsection{Localization performance}

\begin{figure*}[ht]
    \centering
    \subfigure[]{
        \includegraphics[width=\subfigwidth]{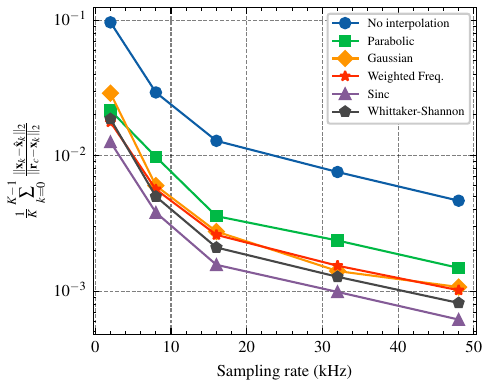}
        \label{fig:tdeinterp2024:sim_fs_pos_crit}}
    \subfigure[]{
        \includegraphics[width=\subfigwidth]{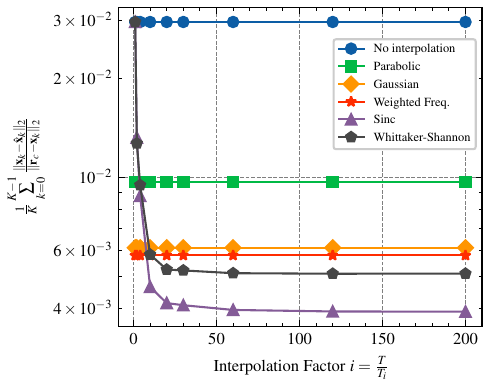}
        \label{fig:tdeinterp2024:sim_interp_pos_crit}} \\
    \subfigure[]{
        \includegraphics[width=\subfigwidth]{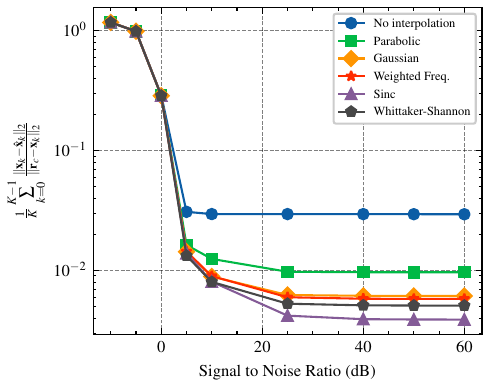}
        \label{fig:tdeinterp2024:sim_snr_pos_crit}}
    \subfigure[]{
        \includegraphics[width=\subfigwidth]{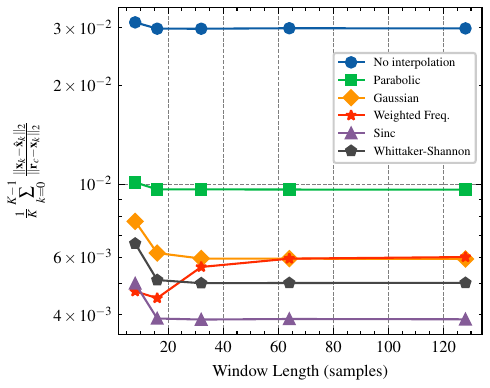}
        \label{fig:tdeinterp2024:sim_frame_len_pos_crit}}
    \caption{The effect of varying (a) the sampling frequency $f_\mathrm{s}$, (b) the interpolation factor $i$, (c) the \gls{snr}, and (d) the window length $L$ on interpolation performance for determining the position in the XY plane of critically sampled acoustic reflections. The non-varying parameters were set to $f_\mathrm{s} = 8$ $\si{\kilo\hertz}$, $i = 200$, \gls{snr} = 40 $\si{\deci\bel}$, $L = 32$ samples, $S_\textrm{sinc} = 1$ sample, and $S_\mathrm{ws} = 9$ samples.}
    \label{fig:tdeinterp2024:sim_pos_crit}
\end{figure*}

\begin{figure*}[ht]
    \centering
    \subfigure[]{
        \includegraphics[width=\subfigwidth]{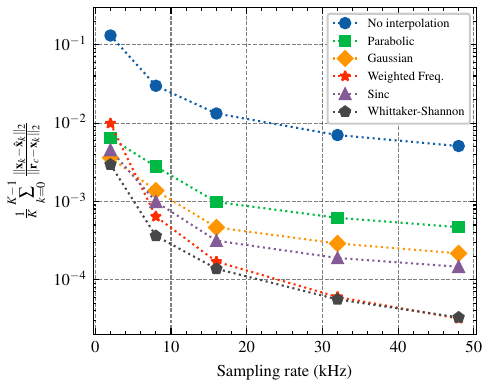}
        \label{fig:tdeinterp2024:sim_fs_pos_band}}
    \subfigure[]{
        \includegraphics[width=\subfigwidth]{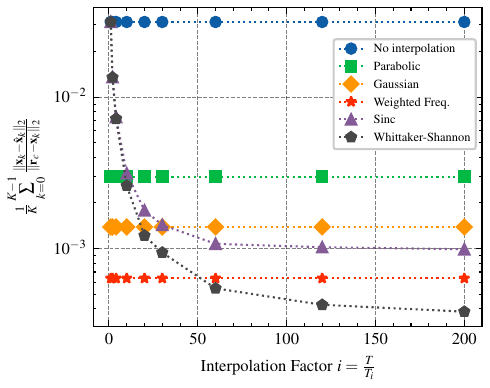}
        \label{fig:tdeinterp2024:sim_interp_pos_band}} \\
    \subfigure[]{
        \includegraphics[width=\subfigwidth]{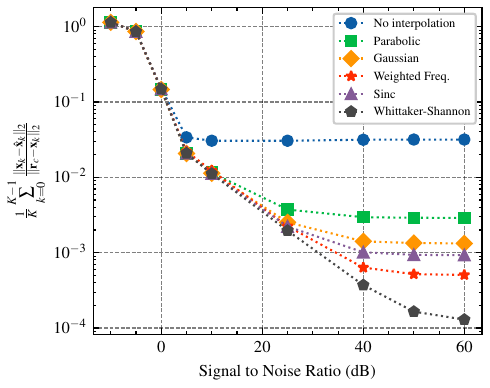}
        \label{fig:tdeinterp2024:sim_snr_pos_band}}
    \subfigure[]{
        \includegraphics[width=\subfigwidth]{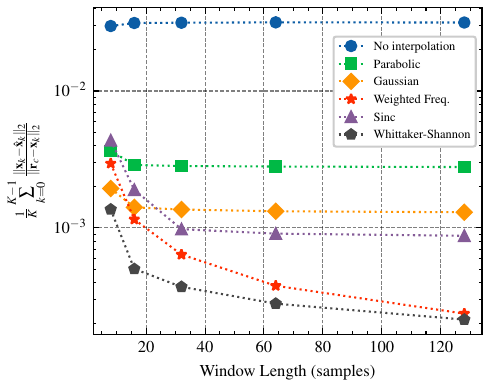}
        \label{fig:tdeinterp2024:sim_frame_len_pos_band}}
    \caption{The effect of varying (a) the sampling frequency $f_\mathrm{s}$, (b) the interpolation factor $i$, (c) the \gls{snr}, and (d) the window length $L$ on interpolation performance for determining the position in the XY plane of band-limited acoustic reflections. The bandlimit was set to $B = \frac{2 f_\mathrm{s}}{5}$. The non-varying parameters were set to $f_\mathrm{s} = 8$ $\si{\kilo\hertz}$, $i = 200$, \gls{snr} = 40 $\si{\deci\bel}$, $L = 32$ samples, $S_\textrm{sinc} = 32$ samples, and $S_\mathrm{ws} = 9$ samples.}
    \label{fig:tdeinterp2024:sim_pos_band}
\end{figure*}

The acoustic source localization performance was assessed by estimating the position for a set of image sources $\mathbf{\hat{x}}_k$ using (\ref{eq:tdeinterp2024:im_position_est}), where the estimated \gls{toa} and \gls{tdoa} values were obtained using the different interpolation methods. The localization performance was evaluated in terms of the mean localization error $\epsilon_\mathrm{pos}$, calculated using a normalized Euclidean distance between the estimated and actual position of the image source. The normalization was done with respect to the distance between the center of the sensor array and the image source. The mean localization error is defined as

\begin{equation}
    \epsilon_\mathrm{pos} = \frac{1}{K} \sum_{k = 0}^{K-1} \frac{\| \mathbf{x}_k - \hat{\mathbf{x}}_k \|_2}{\| \mathbf{r}_c - \mathbf{x}_k \|_2}.
\end{equation}

The results of this analysis can be seen in \Cref{fig:tdeinterp2024:sim_pos_crit} and \Cref{fig:tdeinterp2024:sim_pos_band} for critically sampled and band-limited acoustic reflections, respectively. The results show that the localization performance increases with increasing sampling rates, interpolation factors, and \glspl{snr}. The performance of the interpolation methods for localization is consistent with the results observed for \gls{toa} and \gls{tdoa} estimation. Sinc interpolation achieves the best result for critically sampled reflections, while Whittaker-Shannon interpolation outperforms the other interpolation methods for band-limited acoustic reflections.

A notable observation is that the localization performance is improved for band-limited reflections over critically sampled reflections. This is attributed to the increased number of samples that are available for interpolation around the main lobe of the \gls{tde} function in band-limited conditions. As a result, the interpolation methods can achieve more accurate \gls{toa} and \gls{tdoa} estimates, resulting in improved localization performance. \section{Evaluation with Measured Data}
\label{sec:tdeinterp2024:measurements}

This section assesses the performance of the different interpolation methods discussed in \Cref{sec:tdeinterp2024:interpolation} in achieving acoustic reflector localization with subsample accuracy using real-world experimental measurements. The performance of the interpolation methods is evaluated in terms of the \gls{tdoa} error between the estimated \gls{tdoa} and a ground-truth \gls{tdoa}, and in terms of the positional error of the acoustic reflections.

\subsection{Experimental Setup}
\label{sec:tdeinterp2024:measurements_setup}

The measurements were obtained from the publicly available \textit{MYRiAD} dataset \cite{dietzen2023myriad}, which consists of room acoustic measurements acquired using diverse microphone-array and loudspeaker configurations. The measurements considered in this study were recorded in the \gls{ail}, employing a circular microphone array with a radius of $20$ $\si{\centi\metre}$, comprising eight microphones. A total of $20$ distinct source positions were utilized, resulting in $160$ unique microphone-source pairs. The source positions that were used are denoted as \textit{SL1-SL8} and \textit{SU1-SU12} in the \textit{MYRiAD} dataset. The AIL has a volume of approximately $208$ $\si{\metre\cubed}$ and exhibits a reverberation time of $T_{20} = 0.5$ $\si{\second}$ \cite{dietzen2023myriad}.

The measurements are originally sampled at a rate of $f_\mathrm{s} = 44.1$ $\si{\kilo\hertz}$. To allow down-sampling to $8$ $\si{\kilo\hertz}$ by an integer factor, the measurements were resampled to a sampling rate of $f_\mathrm{s} = 48$ $\si{\kilo\hertz}$.

To compensate for the frequency response of the loudspeakers and microphones, a matched filter $m_n[\kappa]$ was applied to each measured \gls{rir} \cite{antonacci2012inference, diego2021dechorate}. This matched filter was estimated for each loudspeaker-microphone pair by isolating the direct-path component. The matched filter $m_n[\kappa]$ was then correlated with the measured \gls{rir} $\tilde{h}_n[\kappa]$ to obtain the compensated \gls{rir} $h_n[\kappa]$, as,

\begin{equation}
    h_n[\kappa] = \sum_{i = 0}^{M - 1} \tilde{h}_n[\kappa] m_n[\kappa + i],
\end{equation}

\noindent where $M$ is the number of samples in the measured \gls{rir} $h_n[\kappa]$. The compensated \glspl{rir} were used to estimate the \gls{toa} of each acoustic reflection and \glspl{tdoa} between each microphone-source pair.

\begin{figure}[ht]
    \centering
    \includegraphics{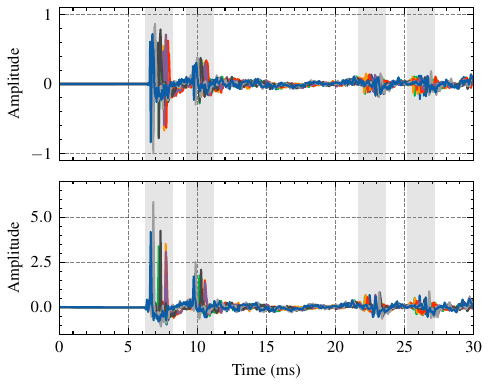}
    \caption{Normalized \glspl{rir} from the \textit{MYRiAD} dataset at loudspeaker position \textit{SL1}, captured using the eight-microphone circular array. The top panel plots the initial 30 $\si{\milli\second}$ of measured \glspl{rir} including the direct path and first three reflections. The bottom panel plots the compensated \glspl{rir} after match-filtering. The gray-shaded regions indicate the time windows used to isolate the direct path and reflections, identified using a peak-picking algorithm.}
    \label{fig:tdeinterp2024:myriad_SL1}
\end{figure}

The direct-path component and first three reflections were subsequently identified by applying a peak-picking algorithm to the match-filtered \glspl{rir}. The acoustic events were isolated using a $2$ $\si{\milli\second}$ long window, corresponding to a window length of $L = 96$ samples. In \Cref{fig:tdeinterp2024:myriad_SL1}, the first $30$ $\si{\milli\second}$ of the \glspl{rir} measured using an eight-microphone circular array at loudspeaker position \textit{SL1} are shown in the top panel, while the first $30$ $\si{\milli\second}$ of the match-filtered \glspl{rir} are shown in the bottom panel. The gray-shaded regions indicate the time windows used to isolate the direct path and reflections, which were identified using the peak-picking algorithm.

The ground-truth \gls{tdoa}, \gls{toa}, and position estimates were obtained at $48$ $\si{\kilo\hertz}$ without applying interpolation methods. These ground truths serve as a reference for evaluating the performance of the different interpolation methods. Limiting the measurements to the direct-path and the first three reflections was done to ensure that the reflections were accurately determined in time and to avoid the influence of multiple reflections in a single time window.

It is worth noting that, due to idealized assumptions made in the derivation of the interpolation methods, model mismatches will inevitably be introduced under real-world measurement conditions. For instance, the specular-reflection assumption made in \Cref{sec:tdeinterp2024:methods} does not hold in practice, since acoustic reflectors impose frequency-dependent filtering on reflected signals. Moreover, the image-source model is based on point-like source and receiver approximations that do not hold in real environments.

The performance of the interpolation methods was evaluated by first down-sampling the isolated early reflections to $8$ $\si{\kilo\hertz}$, resulting in a window length of $16$ samples. Subsequently, the \gls{toa} and \gls{tdoa} were estimated for all microphone-pairs and loudspeaker positions from the downsampled windows while applying the different interpolation methods discussed in \Cref{sec:tdeinterp2024:interpolation}. The performance of all interpolation methods was assessed in terms of the \gls{tdoa} error and positional error, defined as the absolute difference between the estimated \gls{tdoa} and the ground-truth \gls{tdoa}, and the distance between the estimated reflection position and the ground-truth reflection position, respectively. It is important to mention that the results presented in this section are based on the assumption that ground-truth TDOA and TOA estimates are correct, which, due to sampling rate limitations and the presence of noise in the ground-truth measurements, may not be the case in practice. Therefore, the results presented in this section should be interpreted with caution.

The interpolation parameters for sinc interpolation and Whittaker-Shannon interpolation were set to $S_\textrm{sinc} = 3$ samples and $S_\textrm{ws} = 13$ samples, respectively, based on the results obtained in the simulation study presented in \Cref{sec:tdeinterp2024:simulations}. For sinc and Whittaker-Shannon interpolation, the interpolation factor was set to $\mathrm{i} = 500$.

\subsection{Results}
\subsubsection*{\gls{tdoa} Error}

\begin{table}
    \centering
    \caption{Median, mean, and standard deviation of the \gls{tdoa} error for the different interpolation methods on \textit{MYRiAD} dataset measurements. The best performing interpolation method is highlighted in bold.}
    \begin{tabular}{l|ccc}
        \hline
        Interpolation method & Median ($\si{\micro\second}$) & Mean ($\si{\micro\second}$) & Std ($\si{\micro\second}$) \\
        \hline
        No interpolation     & 41.67                         & 31.96                       & 20.67                      \\
        Parabolic            & 7.56                          & 7.55                        & 4.82                       \\
        Gaussian             & 3.97                          & 5.40                        & 4.95                       \\
        Weighted Frequency   & 2.65                          & 4.34                        & 4.85                       \\
        Sinc                 & 2.92                          & 3.46                        & 2.86                       \\
        Whittaker-Shannon    & \textbf{1.67}                 & \textbf{2.30}               & \textbf{2.00}              \\
        \hline
    \end{tabular}
    \label{tab:tdeinterp2024:tdoa_error_myriad}
\end{table}

The \gls{tdoa} error for the different interpolation methods is presented in \Cref{tab:tdeinterp2024:tdoa_error_myriad}. The best performing interpolation method is highlighted in bold. The results show that Whittaker-Shannon interpolation outperforms the other interpolation methods in terms of the median, mean, and standard deviation of the \gls{tdoa} error on the direct-path and first three reflections in the \textit{MYRiAD} dataset measurements.

\begin{figure}[ht]
    \centering
    \includegraphics[width=\subfigwidth]{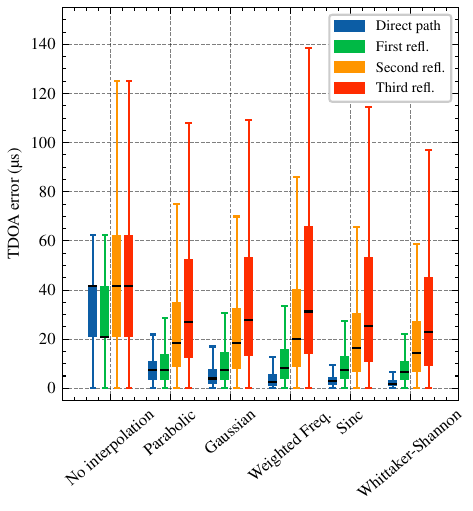}
    \caption{\gls{tdoa} error for the different interpolation methods after isolating the direct-path and first three acoustic reflections in the \textit{MYRiAD} dataset measurements. The results are presented as box plots, where the central mark indicates the median \gls{tdoa} error, the edges of the box represent the $25$th and $75$th percentiles, and the whiskers extend to the most extreme data points excluding outliers.}
    \label{fig:tdeinterp2024:myriad_tdoa_error}
\end{figure}

In \Cref{fig:tdeinterp2024:myriad_tdoa_error}, the \gls{tdoa} error for the different interpolation methods is visualized as box plots for the direct-path and first three reflections in the \textit{MYRiAD} dataset measurements. The results show that Whittaker-Shannon interpolation demonstrates the lowest median \gls{tdoa} error and the smallest interquartile range for the direct-path and first three reflections, indicating that Whittaker-Shannon interpolation provides the most consistent performance across the different microphone-source pairs and loudspeaker positions. Sinc interpolation demonstrates the second-best performance in terms of the median \gls{tdoa} error and the interquartile range. It can also be observed that weighted frequency interpolation performs well for the direct-path and first reflection, but its performance deteriorates for the second and third reflections, as indicated by the larger interquartile range.

\subsubsection*{Positional Error}

\begin{table}
    \centering
    \caption{Median, mean, and standard deviation of the positional error for the different interpolation methods on \textit{MYRiAD} dataset measurements. The best performing interpolation method is highlighted in bold.}
    \begin{tabular}{l|ccc}
        \hline
        Interpolation method & Median ($\si{\centi\metre}$) & Mean ($\si{\centi\metre}$) & Std ($\si{\centi\metre}$) \\
        \hline  
        No interpolation     & 2.49                         & 4.14                       & 3.64                      \\
        Parabolic            & 0.39                         & 0.93                       & 1.09                      \\
        Gaussian             & 0.84                         & 0.95                       & 0.70                      \\
        Weighted Frequency   & 0.57                         & 0.78                       & 0.61                      \\
        Sinc                 & \textbf{0.27}                & 0.56                       & 0.61                      \\
        Whittaker-Shannon    & 0.32                         & \textbf{0.46}              & \textbf{0.42}             \\
        \hline
    \end{tabular}
    \label{tab:tdeinterp2024:pos_error_myriad}
    \vspace{-0.5em}
\end{table}

The positional error for the different interpolation methods is presented in \Cref{tab:tdeinterp2024:pos_error_myriad}. The results show that sinc interpolation demonstrates the lowest median positional error, followed by Whittaker-Shannon interpolation. The results also show that the latter method demonstrates the lowest mean and standard deviation of the positional error.

\begin{figure}[ht]
    \centering
    \includegraphics[width=\subfigwidth]{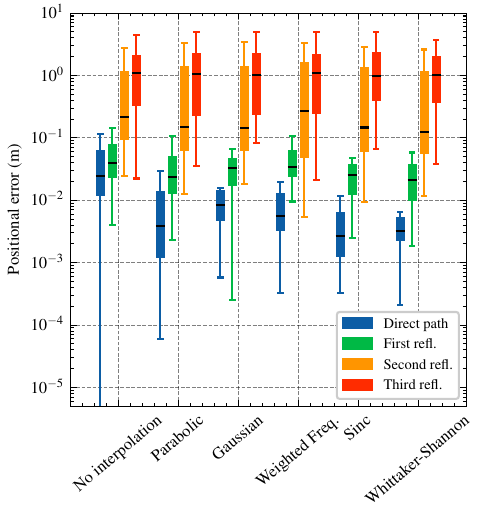}
    \caption{Positional error for the different interpolation methods after isolating the direct-path and first three acoustic reflections in the \textit{MYRiAD} dataset measurements. The results are presented as box plots, where the central mark indicates the median \gls{tdoa} error, the edges of the box represent the $25$th and $75$th percentiles, and the whiskers extend to the most extreme data points excluding outliers.}
    \label{fig:tdeinterp2024:myriad_pos_error}
\end{figure}

In \Cref{fig:tdeinterp2024:myriad_pos_error}, the positional error is visualized for the different interpolation methods as box plots for the direct-path and first three reflections in the \textit{MYRiAD} dataset measurements. The results are plotted on a logarithmic scale and show that sinc interpolation demonstrates the lowest median positional error, followed closely by Whittaker-Shannon interpolation.

The findings of this study indicate that the interpolation methods discussed in this paper can enhance the accuracy of acoustic reflector localization in real-world environments. However, given that the sampling rate utilized in the \textit{MYRiAD} dataset measurements was $44.1 \si{\kilo\hertz}$, the ground truth \gls{tdoa} and reflection position exhibited limited time-resolution. Consequently, the outcomes presented in this section may not fully reflect the performance of the interpolation methods. Future research will focus on evaluating the performance of the interpolation methods in achieving subsample precision in acoustic reflector localization, using real-world measurements with more accurate ground truth estimates.
\vspace{-0.3em} \section{Conclusion}
\label{sec:tdeinterp2024:conclusion}

In this paper, we have presented a comprehensive study on time delay interpolation with subsample accuracy for acoustic reflector localization. We have derived the Whittaker-Shannon interpolation formula from the previously proposed sinc interpolation by assuming infinite support of the sinc function over discrete time. This reformulation allows for the application of Whittaker-Shannon interpolation to windowed \gls{tde} functions, which is essential to achieve accurate estimation in practical scenarios where acoustic reflections are often windowed in reverberant conditions. The Whittaker-Shannon interpolation method has not been previously applied to the subsample \gls{tde} of acoustic reflections. It was shown through simulation that sinc interpolation and Whittaker-Shannon interpolation are able to outperform parabolic, Gaussian, and weighted frequency interpolation in terms of the \gls{tdoa} error and positional error for critically sampled and band-limited signals, respectively. The simulations also show that sinc and Whittaker-Shannon interpolation is robust for small window lengths, low sampling rates, and low \glspl{snr} compared to existing interpolation methods. The performance of the interpolation methods was further evaluated on real-world measurements from the \textit{MYRiAD} dataset, showing that sinc and Whittaker-Shannon interpolation are able to outperform existing interpolation methods in terms of the \gls{tdoa} error and positional error for the direct-path and first three reflections. The results indicate that Whittaker-Shannon interpolation provides the most consistent performance across different sensor-source pairs and loudspeaker positions. The results of this study can be used to improve the accuracy of acoustic reflector localization methods, which are essential for various applications, such as room acoustics analysis, sound source localization, and acoustic scene analysis. 
\begin{acknowledgments}
This research work was carried out at the ESAT Laboratory of KU Leuven, in the frame of FWO Large-scale research infrastructure "The Library of Voices - Unlocking the Alamire Foundation's Music Heritage Resources Collection through Visual and Sound Technology" (I013218N), FWO SBO Project "The sound of music - Innovative research and valorization of plainchant through digital technology" (S005319N), and SBO Project "New Perspectives on Medieval and Renaissance Courtly Song" (S005525N). This research received funding from the Flemish Government (AI Research Program). This research received funding from internal KU Leuven funds: C14/21/075 "A holistic approach to the design of integrated and distributed digital signal processing algorithms for audio and speech communication devices”, C3/23/056 "HELIXON: Hybride, efficiënte en vloeiende interpolatie van geluid in uitgebreide realiteit”. The research leading to these results has received funding from the European Research Council under the European Union's Horizon 2020 research and innovation program / ERC Consolidator Grant: SONORA (no. 773268). This paper reflects only the authors' views and the Union is not liable for any use that may be made of the contained information.
\end{acknowledgments}

\section*{Author Declarations}

The authors have no conflicts to disclose.

\section*{Data Availability}

The code used to generate the simulation results presented in this paper is available at \dourl{https://github.com/hrosseel/TDE_INTERP}.

\end{document}